\documentstyle[12pt]{article}

\newcommand{\mpl}{M_{\rm pl}}
\newcommand{\mst}{M_{*}}

\newcommand{\be}{\begin{equation}}
\newcommand{\ee}{\end{equation}}
\newcommand{\bea}{\begin{eqnarray}}
\newcommand{\eea}{\end{eqnarray}}

\newcommand{\newc}{\newcommand}
\newc{\gsim}{\lower.7ex\hbox{$\;\stackrel{\textstyle>}{\sim}\;$}}
\newc{\lsim}{\lower.7ex\hbox{$\;\stackrel{\textstyle<}{\sim}\;$}}
\newc{\gev}{\,{\rm GeV}}
\newc{\mev}{\,{\rm MeV}}
\newc{\ev}{\,{\rm eV}}
\newc{\kev}{\,{\rm keV}}
\newc{\tev}{\,{\rm TeV}}

\renewcommand{\phi}{\varphi}
\newc{\smu}{{\tilde\mu}}
\newc{\snu}{{\tilde\nu}}
\newc\order{{\cal O}}
\newc{\eps}{\epsilon}
\newc{\re}{\mbox{Re}\,}
\newc{\im}{\mbox{Im}\,}
\newc{\lunits}{\,\mbox{cm}^{-2}\mbox{s}^{-1}}
\relax
\def\boxeqn#1{\vcenter{\vbox{\hrule\hbox{\vrule\kern3pt\vbox{\kern3pt
\hbox{${\displaystyle #1}$}\kern3pt}\kern3pt\vrule}\hrule}}}
\def\qed#1#2{\vcenter{\hrule \hbox{\vrule height#2in
\kern#1in \vrule} \hrule}}
\newc{\ie}{{\it i.e.}}          \newc{\etal}{{\it et al.}}
\newc{\eg}{{\it e.g.}}          \newc{\etc}{{\it etc.}}
\newc{\cf}{{\it c.f.}}
   \def\CL{{\cal L}}
  \def\CO{{\cal O}}

\def\ltap{\ \raise.3ex\hbox{$<$\kern-.75em\lower1ex\hbox{$\sim$}}\ }
\def\gtap{\ \raise.3ex\hbox{$>$\kern-.75em\lower1ex\hbox{$\sim$}}\ }
\def\gl{\ \raise.5ex\hbox{$>$}\kern-.8em\lower.5ex\hbox{$<$}\ }
\def\roughly#1{\raise.3ex\hbox{$#1$\kern-.75em\lower1ex\hbox{$\sim$}}}
\def\dsl{\,\raise.15ex\hbox{/}\mkern-13.5mu D} 
\def\delsl{\raise.15ex\hbox{/}\kern-.57em\partial}
\def\Ksl{\hbox{/\kern-.6000em\rm K}}
\def\Asl{\hbox{/\kern-.6500em \rm A}}
\def\Dsl{\hbox{/\kern-.6000em\rm D}} 
\def\Qsl{\hbox{/\kern-.6000em\rm Q}}
\def\gradsl{\hbox{/\kern-.6500em$\nabla$}}
\let\beta=\beta

\let\Ga=\Gamma
\let\de=\delta
\let\De=\Delta

\newdimen\pmboffset
\def\oldpmb#1{\setbox0=\hbox{#1}%
 \copy0\kern-\wd0
 \kern\pmboffset\raise 1.732\pmboffset\copy0\kern-\wd0
 \kern\pmboffset\box0}

\def\bar#1{\overline{#1}}
\def\vev#1{\left\langle #1 \right\rangle}

\def\inv{^{\raise.15ex\hbox{${\scriptscriptstyle -}$}\kern-.05em 1}}
\def\pr#1{#1^\prime}  
\def\lbar{{\lower.35ex\hbox{$\mathchar'26$}\mkern-10mu\lambda}} 

\let\<=\langle
\let\>=\rangle

\let\+=\uparrow
\let\-=\downarrow

\begin{document}

\baselineskip=17pt
\pagestyle{plain}
\setcounter{page}{1}

\begin{titlepage}

\begin{flushright}
CERN-TH/99-38\\
SLAC-PUB-8068\\
SU-ITP-98/68

\end{flushright}
\vspace{10 mm}

\begin{center}
{\LARGE Rapid Asymmetric Inflation and Early Cosmology}
\vskip 2mm
{\LARGE in Theories with Sub-Millimeter Dimensions }
\vspace{3mm}
\end{center}
\begin{center}
{\large Nima Arkani-Hamed$^a$, Savas Dimopoulos$^b$,
Nemanja Kaloper$^b$,}\\
{\large and John March-Russell$^c$}\\ \vspace{3mm} {\em
$^a$ SLAC, Stanford University, Stanford, CA 94309, USA}\\
{\em $^b$
Physics Department, Stanford University, Stanford CA 94305, USA}\\
{\em $^c$ Theory Division, CERN, CH-1211, Geneva 23, Switzerland}
\end{center}
\vspace{3mm}
\begin{center}
{\large Abstract}
\end{center}
\noindent
It was recently pointed out that the fundamental Planck mass
could be close to the TeV scale with the observed weakness of
gravity at long distances being due the
existence of new sub-millimeter spatial dimensions.
In this picture the standard model fields are
localized to a $(3+1)$-dimensional wall or ``3-brane''.
We show that in such theories there exist attractive models
of inflation that occur while the size of the new dimensions
are still small.  We show that it is easy to produce the
required number of efoldings, and further that the density perturbations
$\delta\rho/\rho$ as measured by COBE can be easily reproduced, both
in overall magnitude and in their approximately scale-invariant spectrum.
In the minimal approach, the inflaton field is just the
moduli describing the size of the internal dimensions, the role
of the inflationary potential being played by the
stabilizing potential of the internal space.
We show that under quite general conditions, the inflationary
era is followed by an epoch of contraction of our world on
the brane, while the internal dimensions slowly expand to their
stabilization radius.  We find a set of exact solutions which
describe this behavior, generalizing the well-known Kasner solutions.
During this phase, the production of bulk gravitons remains suppressed.
The period of contraction is terminated by the blue-shifting of Hawking
radiation left on our wall at the end of the inflationary de~Sitter phase.
The temperature to which this is reheated is consistent with
the normalcy bounds.  We give a precise definition of the radion moduli
problem.
\end{titlepage}
\newpage


\section{Introduction}

It was recently pointed out that the fundamental Planck mass
could be close to the TeV scale\cite{ADD,AADD,ADDlong,AHDMR}, thus
providing a novel solution to the hierarchy problem for the standard
model.  Gravity becomes comparable in
strength to the other interactions at this scale, and the
observed weakness of gravity at long distances is then explained
by the presence of $n$ new ``large" spatial dimensions.
Gauss' Law relates the Planck scales of the
$(4+n)$ dimensional theory, $\mst$, and the long-distance
4-dimensional theory, $\mpl$,
\be
\mpl^2 =(b_0)^n \mst^{n+2}
\label{gauss}
\ee
where $b_0$ is the (present, stabilized) size of the extra
dimensions.  If we put $\mst \sim 1\tev$ then
\be
b_0 \sim 10^{-17+\frac{30}{n}} \mbox{cm}
\ee
For $n=1$, $b_0 \sim 10^{13}$ cm, so this case is
excluded since it would modify Newtonian gravitation at
solar-system distances. Already for $n=2$, however, $b_0 \sim 1$
mm, which happens to be the distance where our present experimental
knowledge of gravitational strength forces ends.  For larger
$n$, $1/b_0$ slowly approaches the fundamental Planck scale
$\mst$.

While the gravitational force has not been measured beneath
a millimeter, the success of the SM up to $\sim 100\gev$ implies
that the SM fields can not feel these extra large dimensions; that
is, they must be stuck on a 3-dimensional wall, or ``3-brane", in the
higher dimensional space.  Thus, in this framework the universe is
$(4+n)$-dimensional with fundamental Planck scale near the weak
scale, with $n \geq 2$ new sub-mm sized dimensions where gravity,
and perhaps other fields, can freely propagate, but where the SM
particles are localized on a 3-brane in the higher-dimensional space.
The most attractive possibility for localizing the SM fields to
the brane is to employ the D-branes that naturally occur in type I
or type II string theory \cite{Dbrane,AADD}.  Gauge and other
degrees of freedom are naturally confined to such
D-branes~\cite{Dbrane}, and furthermore this approach has the obvious
advantage of being formulated within a consistent theory of gravity.
However, from a practical point of view, the most important question
is whether this framework is experimentally excluded.  This was the
subject of \cite{ADDlong} where laboratory, astrophysical, and
cosmological constraints were studied and found not to exclude these
ideas.

There are a number of other papers discussing related suggestions.
Refs.~\cite{DDG} examine the idea of lowering the gauge-coupling
unification scale by utilizing higher dimensions.  Further papers
concern themselves with the construction of string models with extra
dimensions larger than the string scale \cite{antoniadis,HW,tye},
gauge coupling unification in higher dimensions
without lowering the unification scale~\cite{bachas},
the effective theory of the low energy degrees of freedom in realizations of our world
as a brane \cite{Raman1} and radius stabilization \cite{Raman2,AHDMR}.  
There have also been many recent papers
discussing various theoretical and phenomenological aspects of this
scheme \cite{new}, and a few papers on aspects of the early universe
cosmology \cite{bendav}\cite{lyth}\cite{kl}\cite{dt}\cite{mazumdar}
that discuss issues related to those considered here.\footnote{
However, we note that a closer scrutiny of \cite{mazumdar}
reveals that it is impossible to get any inflation at
all in the models considered. The potentials
in \cite{mazumdar} all have exponential
dependence on the radion field, and thus lead to
the power-law dependence of the scale factor
on time, $\bar a \sim t^{2/\beta^2}$ \cite{bdinf}. The correct
formula for the parameter $\beta$ of \cite{mazumdar}
is $\beta = 2\sqrt{2n/(n+2)}$, and thus $\beta \ge 2$
for all cases $n \ge 2$, which gives only subluminal expansion,
and not inflation. In fact, a complete
analysis must employ the physically meaningful measure of scales, namely
the scale factor expressed in units of the Compton wavelength of
wall particles.  (Because of the possibility of Weyl rescaling between
string and Einstein frames one must be careful about the definition
of physical quantities.)  In such physical units it is
possible to show that the scale factor
always expands sub-luminally, even more slowly than the naive
Einstein frame expansion, and thus
none of the solutions of \cite{mazumdar} contain a stage of
inflation.}

In this paper we will discuss inflation and other general
aspects of early universe cosmology in world-as-a-brane models.
In particular we will be concerned with aspects of early
universe cosmology that involve the dynamics of the
internal dimensions in a central way.
We find that there exist attractive models of inflation that occur while
the internal dimensions are still small, far away from their
final stabilized value given by the Gauss' law constraint.  We show
that in such models it is very easy to produce the required number of
efoldings of inflation even though no energy
densities exceed the fundamental Planck scale, $\rho<\mst^4$,
and that supersymmetry, if it exists at all, is very badly
broken on our brane.  Further, we demonstrate that the density
perturbations $\delta\rho/\rho$ as measured by COBE and the other
microwave background and large-scale structure experiments can be
easily reproduced in such models, both in their magnitude and in
their approximately scale-invariant spectrum.
In the most minimal approach, the inflaton field may be just the
moduli describing the size of the internal dimensions itself
(the radion field of \cite{AHDMR}), the role
of the inflationary potential being played by the
stabilizing potential of the internal space.
In the case of a wall-localized inflaton, the cosmological
constant might even result from the electroweak phase
transition, in which case the inflaton is the Higgs.
Actually, an important remark in this regard is that when
the internal dimensions are small, $b \sim \mst^{-1}$, the
distinction between on-the-wall and off-the-wall physics
is not meaningful: \eg, the inflationary features in $V(b)$
at small $b$ could be {\it due} to Higgs physics on the wall.

The approximately scale-invariant nature of the primordial
perturbation spectrum implies that,
during inflation, the internal dimensions must expand more slowly than
the universe on the wall.  Thus we are led to consider
a form of asymmetric inflationary expansion of the
higher-dimensional world.
It is very interesting to note that requiring
the consistency and naturalness of both the long duration of
inflation (\ie, number of efolds $N_e\gtap 100$), and the
magnitude of $\de\rho/\rho$, implies that inflation should
occur early, with the internal dimensions close to their
natural initial size $b_i \sim \mst^{-1}$.
In particular, we avoid the introduction of extremely light
inflatons which seem to be needed if inflation
occurs {\it after} the internal dimensions reach their
current, large, size \cite{bendav,lyth}.  We emphasize here
that this constraint emerged from the assumption that during
inflation the internal space was already large and stable.
In fact, in such scenarios obtaining $\de\rho/\rho$
requires that the inflaton is even lighter \cite{kl}, and
further, that inflation occurring only after stabilization
cannot explain the age of the universe \cite{kl}.
Indeed, the wall-only inflation cannot begin
before $t \sim H^{-1} \sim \mpl/\mst^2 >> \mst^{-1}$,
when the universe is already very large and old.

However, in the models of inflation at very early times
which we consider, the effective $4D$ Planck mass is both
much smaller than it is now, and in general, time-dependent
due to the variation in the volume of the internal dimensions.
Moreover, inflation occurs soon after the ``birth'' of the universe
when the sizes of all dimensions are close to their natural
initial size $\sim \mst^{-1}$.
Hence early asymmetric inflation solves the age problem too.

The framework in which we discuss these issues is that
of semiclassical $(4+n)$-dimensional gravity with an
additional potential $V(b)$ depending on the size of the
internal dimensions.  As discussed in Ref.\cite{AHDMR} this
rapidly becomes a good approximation at energy scales below
$\mst$, which, self consistently, is the correct regime for
early inflation, essentially because of the COBE constraint
that the density perturbations are small
$\de\rho/\rho \sim 2 \times 10^{-5}$.  Most of our analysis
will utilize the so-called string frame with explicit scale
factors $(a(t),b(t))$ for the sizes of our brane-localized
dimensions and the internal dimensions respectively.  In this
frame the effective long-distance 4-dimensional value of Newton's
constant depends on $b(t)$ and changes with time.  However this
is the frame where the ``measuring sticks'' of particle masses
and Compton wavelengths are fixed.
On the other hand,
in the analysis that follows we will find it occasionally
useful to employ the correspondence of the low-energy
$(4+n)$-dimensional theory with an effective 4-dimensional theory
which takes the form of a scalar-tensor theory of gravity, the radius
of the large internal space $b$ playing a role similar
to the Brans-Dicke scalar field.  The inflationary dynamics
from the point of view of this frame is completely equivalent to
the usual slow-roll scenarios, and the conditions for the asymmetric
expansion of the universe on the brane relative to the internal
dimensions are completely equivalent to the slow roll conditions in
the usual inflationary models.  Further, this analogy provides very
useful consistency check for the determination of the density
perturbations.  The relationship between the scale factor $b(t)$
and the canonically normalized 4-dimensional (``radion'') field
is given in equation (\ref{bcanonical}).  (We will often
abuse our terminology and refer to $b(t)$ itself as the
``radion''.)

It is important to realize that under quite general conditions,
the early inflationary era is followed by a long
epoch where the scale factor of our brane-universe undergoes
a slow {\it contraction} while the internal dimensions continue
to expand towards their final stabilized value.
We show that even with the inclusion of a potential for $b$, it
is possible to exactly describe the evolution during this epoch, and
we present a class of exact solutions which generalize the usual
vacuum Kasner solutions.  The total amount of
contraction of our universe on the brane is bounded
by a small power of the expansion of the size of internal
dimensions, and varies between at most $7$ efoldings in the case
of two extra dimensions to at most $12$ efoldings when there are
six extra dimensions.
We show that during this phase of $b(t)$ evolution to the
stabilization point, the production of bulk
gravitons by the time-varying metric remains completely suppressed,
ensuring that the bulk is very cold at, and after, the stabilization
of the internal dimensions.

However, the particles produced by gravitational (Hawking)
effects on the brane at the end of the inflationary de~Sitter
phase can play an important role.  Their energy density
is blue-shifted by the slow contraction of $a(t)$, and when
this wall-localized energy density exceeds the energy density
in the radion, the contraction of $a(t)$ ceases, with a ``Big Bounce''
occurring, leading on to an expansion of both $a$ and $b$.
It is very tempting to view the reheating on the wall as due to
this contraction of primordial de~Sitter radiation.  This
is especially so as the final temperatures
come naturally close to the normalcy bounds (the upper bound
$T_*$ on the temperature of \cite{ADDlong} that ensure that
the evaporation to bulk gravitons remains negligible).

However, such a conclusion is premature, since as we argue
below the energy density in the radion is naturally of the same
order as that on the wall, and this leads to a radion moduli
problem in the later evolution of the universe.  Explicitly,
because of the gravitationally suppressed couplings of this
field to wall-localized SM states, the radion decays
very slowly back into light wall fields.
As a result, its energy density would red-shift only
according to $\rho_b \sim 1/a^3$ compared to
$\rho_{wall} \sim 1/a^4$, and comes to dominate
the total energy density.  This is just the standard
moduli problem, which is inextricably linked with
the mechanism of reheating in world-as-a-brane models.

Concretely, the basic picture we advocate is:
\begin{itemize}
\item
The quantum creation of the universe takes place with the initial
size of all dimensions close to the fundamental Planck scale
$\mst^{-1}$.  In particular the initial size $b_i$ of the internal
dimensions is of this size.
\item
A prolonged period of inflation in a direction parallel to
our brane takes place, with our scale factor $a(t)$
increasing superluminally $a(t)\sim t^p$ with
$p>>1$, and $b(t)=b_I$ essentially static. (In
the cases that we explicitly discuss the inflation is
quasi-exponential, $a(t)\sim a_i \exp(H t)$.)  This
brane inflation is driven by
either the stabilizing potential of the radion $b$ itself, or by
a wall-localized field with effective non-zero cosmological constant.
In either case it is unnatural for the size of the effective
4-dimensional cosmological constant to exceed roughly $(\tev)^4$.
Since the internal dimensions are small the effective
4-dimensional Newton's constant is large
\be
G_{N,initial} = {1\over b_I^n \mst^{n+2}} \simeq {1\over \mst^2}.
\label{initialG}
\ee
Thus the Hubble constant during this initial period of inflation
can be {\it large} even though the energy density is quite small,
$\vev{V}\sim O(\tev^4)$,
\be
H_{\rm infl}^2 \simeq {\vev{V}\over b_I^n\mst^{n+2}} \simeq {\vev{V}\over
\mst^2}.
\ee
Thus inflation can be {\it rapid}, and moreover, as we will
argue in detail the density
perturbations can be large, being determined in order of magnitude
to be
\be
{\de\rho\over \rho} \simeq {H_{\rm infl}\over \mst (\mst b_I)^{n/2} S}.
\ee
where $S$ is a parameter that encapsulates both the duration
of $a(t)$ inflation and the deviation of the
perturbation spectrum from
the scale-invariant Harrison-Zeldovich spectrum.
(We will argue that $S\ltap 1/50$.)
We can turn this around by imposing the COBE-derived normalization
on $\delta\rho/\rho$, and thus on $\vev{V}/ b_I^n\mst^{n+2}$.
During this period the size $b_I$ of the internal radii are fixed
by the over-damping arising from $H_{\rm infl}$.
\item
Wall inflation now ends, with $H$ starting to drop and simultaneously
the radion starting to evolve to its' minimum at $b_0$.  The initial
motion of the radion can be quite involved, but from the coupled
equations of motion for the scale factors $a(t)$ and $b(t)$
we will see that as $b$ evolves towards its' minimum, our scale
factor, under very general conditions, undergoes a collapse.
The detailed behavior generalizes the well-known Kasner solutions
where some dimensions expand, while others collapse, both with
determined (subluminal) power-law dependence on time.
\item
The contraction epoch ends when the blue-shifting radiation
density on our brane becomes equal to the energy density in the
radion field $\rho_{wall} = \rho_b$.  A model-independent
initial source for the radiation on the wall is the Hawking
radiation left over from the early inflationary de~Sitter
phase.  After the contraction of $a(t)$ reverses, the
radion and wall-localized energy densities scale together
until the stabilization point is reached.
\item
Finally around the stabilization point $b_0$ the radion field
starts to oscillate freely.  Since this energy density scales as
$1/a^3$, and the wall-to-radion energy densities are initially
comparable at the start of the oscillation era, the radion
energy starts to dominate the total energy density.
\item
The most serious question that early universe cosmology presents
in the world-as-a-brane scenario is how do we dilute this energy
in radion oscillations to an acceptable level.  The radion is
long-lived, its' decay width back to light wall states being given
by\footnote{Here $m_\phi$ is the mass of the
canonically normalized field corresponding to
$b$.  The experimental bound on this mass is $m_\phi \gtap 10^{-3}\ev$.
Note that the decay width is increased if we take into account the possible
presence of many branes in the bulk.  Indeed some scenarios of radius
stabilization \eg, the ``brane crystallization" picture
\cite{AHDMR}, require
$N_{\rm wall} \simeq (\mpl/\mst)^{2(n-2)/n}$ branes in the bulk.
If each of these have $O(1)$ light modes then the total decay
width to all branes is greatly enhanced.}
\be
\Ga_{\phi} \simeq {m_\phi^3\over \mpl^2}.
\ee
We thus require some dilution in the radion
energy density, either by a short period of late
inflation followed by reheating, or by a delayed reheating
after $\rho_{b}$ has sufficiently red-shifted.
The amount of dilution of the radion energy density
that we require is given roughly by $1\ev/T_*\sim 10^{-7}$,
so that only about 5 efolds of late inflation is needed.
\end{itemize}

The paper is organized as follows. In Section~2 we will discuss
the details of the inflationary stage, both in the original
string (brane) frame and in the effective Einstein frame,
showing the initial conditions, the conditions on inflationary
potentials and the resolution of the COBE constraints.
Section~3 is devoted to the study of the era of radion evolution
to stabilization.  In subsection~3.1 we derive the exact solutions which
generalize simple vacuum Kasner solutions, and which
demonstrate that a period of $a(t)$ contraction subsequent to
inflation is quite generic. In subsection~3.2 we discuss
the phenomenological consequences of this era, including
the cause of its cessation and the (minimal) production of
bulk gravitons.  There we will also show how the normalcy
bounds for the temperature of our brane after stabilization
are met by the evolution.  In subsection~3.3 we discuss we
discuss the issue of the production of bulk gravitons during
the era of radion evolution and show that the quantities produced
are harmless.  In subsection~3.4 we provide a physically quite useful
discussion of the contraction and stabilization eras in the Einstein
frame, where many of the arguments of the previous subsections can be
understood quite simply.  In subsection~3.5 we provide
a short discussion of the radion
moduli problem, and quantify its' size using the evolution
equations discussed earlier in Section~3.  Section~4
gives our conclusions.  We also include three extensive Appendices:
Appendix~A contains a short discussion of some of the basic
kinematics of brane evolution embedded in higher dimensions.
Appendix~B gives some additional details of the exact solutions
for the post-inflation slow evolution era where $a(t)$ contracts.
Appendix~C proves that the contraction of $a(t)$ is self-ceasing by the
blue-shifting of brane-localized radiation, and also discusses the
exact solutions that may be obtained for the ``Big Bounce'' that
terminates this contraction.

\section{Early inflation}

We now embark on our detailed discussion of the evolution
of the sizes of our brane and the transverse ``internal'' dimensions.
In particular in this section we focus on the physics of the
(early) inflationary epoch.

We are interested in the case of a 3-brane embedded in a
$(4+n)$-dimensional
spacetime.  To reproduce our observed world, ultimately the 3
spatial dimensions parallel to the brane must be as least as large
as our current horizon size, while the $n$ transverse spatial dimensions
have to stabilize at a size $b_0$ given by the constraint
on $G_N$, ({\ref{gauss}}),
$(b_0)^n \mst^{n+2} = \mpl^2$.  Note that we take the internal space to be
topologically compact from the beginning.
That is we impose periodic boundary conditions in
the directions transverse to the wall. These conditions reflect
the fact that the low energy theory is four-dimensional, and that
the classical evolution preserves the topology.

The total action is comprised of a bulk part,
\be
S_{\rm bulk}  = - \int d^{4+n}x \sqrt{-\det G_{(4+n)}} \biggl(
\mst^{(n+2)}{\cal R} - {\cal L}_{\rm matter} + \ldots
\biggr),
\label{bulkS}
\ee
and a brane part,
\be
S_{\rm brane}  =  - \int d^{4}x \sqrt{-\det g_{(4)}^{\rm induced}}
\biggl({\cal L}_{\rm standard~model} + \ldots \biggr),
\label{actions}
\ee
where ${\cal L}_{\rm matter}$ is the Lagrangian of the bulk
fields apart from the graviton.  These fields give rise to
the stabilizing potential $V$ that we discuss below.  The
ellipses denote higher-derivative terms that can
be safely ignored in most of the regime of interest
as curvatures are small compared to the fundamental Planck
scale $\mst$, apart possibly from the very early pre-inflationary
stage immediately after the quantum creation of the universe.
However any signatures of such a phase of high energy and
large curvature have been wiped out of the visible
universe by the subsequent stage of inflation.
We will therefore ignore this stage as practically invisible,
and begin the description of the world-as-a-brane
universe at curvatures an order of magnitude or so below the
fundamental Planck scale.  Indeed, at and below these scales,
the description based on the actions (\ref{bulkS}) and
(\ref{actions}) should be reliable.
The background metric for the $(4+n)$-dimensional spacetime
which is consistent with the symmetries of the brane-bulk system
is of the form
\be
g_{\mu\nu} = \left(
\begin{array}{ccc}
1 & & \\
   & - a(t)^2 g_{IJ} & \\
        & & -b(t)^2 g_{ij}
\end{array}
\right) ,
\label{metricform}
\ee
where $a$ is the scale factor of the 3-dimensional space, and
$b$ is the scale factor of the internal $n$-dimensional space,
with geometry set by $g_{ij}$ where $\det(g_{ij}) = 1$.

As shown in Appendix~A, the equations of motion for the coupled
$\{ a(t),b(t)\}$ system can be written in the form,
\be
\dot H_a + n H_a H_b + 3 H_a^2 = {1\over 2(n+2)b^n \mst^{(n+2)}}
\left( b {\partial V\over \partial b} - (n-2) V\right)
\label{aeom}
\ee
and
\be
\dot H_b + 3 H_a H_b + n H_b^2 = -{1\over (n+2)b^n \mst^{(n+2)}}
\left( \frac{b}{n} {\partial V\over \partial b} - 2 V\right)
\label{beom}
\ee
where we have introduced the Hubble parameters $H_a\equiv \dot a /a$ and
$H_b \equiv \dot b / b$ for the two scale factors, and an overdot
denotes a derivative with respect to $t$.
These equations of motion are supplemented by the constraint equation
\be
6 H_a^2 + 6n H_a H_b + n(n-1) H_b^2 = {V\over b^n \mst^{(n+2)}}.
\label{constraint}
\ee
Note that in these formulae the potential $V$ is the effective
4-{\it dimensional} potential with mass dimension $[V]=4$ --
\ie, that arising from projecting the bulk energy density that
results from $\CL_{\rm matter}$ onto our wall.

The natural inflationary initial conditions consist of
taking the initial $(4+n)$-dimensional universe to be relatively smooth,
flat and potential energy-dominated over scales roughly given
by the fundamental
Planck length $1/\mst$, and not much greater~\cite{andc,book}.
In the brane language the initial conditions we adopt
mean that we take a portion of the brane of
linear size $\sim 1/\mst$ embedded in a $(4+n)$-dimensional volume of
the same linear size, to be relatively flat, smooth and straight.
Indeed, it is most natural to suppose that the universe starts as a
small $(4+n)$-dimensional domain of linear extent $1/\mst$
in all directions, and then undergoes an
epoch of inflationary expansion almost immediately, at least
in the directions parallel to our brane.

Quantitatively these conditions can be stated in terms
of the initial conditions for the brane and bulk horizons as
follows. The size of initially isotropic
and homogeneous causal domains must be
\be
H^{-1}_{a,i} \sim H^{-1}_{b,i} \ge \mst^{-1}
\label{initialhubble}
\ee
Moreover we can also define the initial conditions for the
scale factor of the brane $a$ and the radion field $b$ by
referring to the flatness problem.  Since close to the
fundamental Planck scale the energy density on the wall and
in the bulk is of order unity in fundamental Planck units,
if we pick the gauge such that any intrinsic spatial curvature
on the brane or in the bulk is $k=\pm 1$, it cannot exceed
in magnitude $\rho_p$, leading to the conditions that\footnote{In
spatially flat FRW universes
the magnitude of the scale factor $a$ is physically
meaningless. However, the spatial curvature, namely the
quantity $k/a^2$ is physical.  Thus choosing the value
of $a$ at some instant $t$ corresponds to specifying curvature
of spatial hypersurfaces.  Equivalently, choosing the value
of the constant $k$, by normalizing it to $\pm 1$,
corresponds to picking the units for the scale factors. We will
assume this throughout this work, even if we do not
explicitly specify the spatial curvatures (which can be ignored
after prolonged inflation).}
\be
a_i \sim b_i \ge \mst^{-1}
\label{initialsize}
\ee
It is easy to recognize the conditions
(\ref{initialhubble}) and (\ref{initialsize}) as precisely the
consistent initial conditions for inflation, in the case
when the fundamental Planck scale is $M_*$. However these
initial conditions are merely a rough estimate coming from
the requirement that the approximation based on semiclassical
gravity is valid, and that the usual cosmological problems (horizon,
flatness, homogeneity \etc.) are solved by subsequent
inflation.

Further, the generation of sufficient density perturbations
$\delta \rho/\rho\sim \delta \rho/\rho|_{\rm COBE}\sim 10^{-5}$,
implies, at least without the unappetizing introduction of very
light fields on our wall, or
of the assumption of very unusual initial conditions,
that the inflationary epoch that
solves the flatness and horizon problems occurs when the size of
the internal dimensions is still relatively small.  The basic reason
for this is that in a theory with fundamental Planck scale $\mst$,
the energy density localized on the brane should not exceed
$\mst^4$.  Similarly the energy density in the bulk should
not exceed $\mst^{n+4}$.  Moreover, the (matter) energy density of
the universe at its' birth is expected to be of order $\mst^{n+4}$.
If inflation occurs when the internal dimensions
are already large, then the effective Newton's constant in
our 4-dimensions is already very small
$G_{\rm N, eff} = 1/(\mst^{n+2} b^n) \ll 1/\mst^2$, and energy
densities of order $\mst^4$ or less will lead to a very small
Hubble constant, and thus typically unacceptably small density
perturbations.\footnote{Note though that if the internal dimensions
are large $b\gg 1/\mst$ a small bulk $(4+n)$-dimensional energy
density still allows the effective ``projected'' energy
density on the wall to exceed $\mst^4$, without requiring the use of
the full quantum theory of gravity; the semiclassical approximation
is still good.}

To see this in detail consider the expression for the density perturbations
generated during slow-roll inflation driven by a (canonically
normalized) field $\phi$
\be
{\de\rho\over \rho} = \frac{5}{12\pi}{H_{\rm infl}^2\over \dot\phi}.
\label{drho}
\ee
(The use of this 4-dimensional expression will be
self-consistently justified later on in our analysis.)
The Hubble constant on our brane during inflation is given in order
of magnitude by $H_{\rm infl} \simeq V/(\mst^{n+2} b_I^n)$, where
$b_I$ denotes the size of the internal dimensions during the inflationary
epoch.  Since $V\ltap \mst^4$ if $b_I\mst\gg 1$ then ${\de\rho/ \rho}$ is
very small unless $\dot\phi$ is extremely small.  Although this is a
logical possibility, it requires extraordinary fine-tuning and we will
not consider this case here.  On the other hand if $b_I\mst$ is
$\CO(1)$ then sufficiently large density perturbations easily
result.  Moreover, from the power dependence of $H_{\rm infl}$
on $b_I$ we see that {\it if the size of the internal dimensions
changes significantly during inflation the spectrum of density perturbations
will be very far from scale invariant.}  Since the spectral index of
density perturbations $n_\rho$ is constrained by the
cosmic microwave background (CMB) and
large scale structure measurements to be not significantly different
from the scale-invariant value $n_\rho=1$,
\be
|n_\rho-1| < 0.2,
\label{nrho}
\ee
it is necessary that the evolution of the internal dimensions
is slow compared to that of our scale factor $a$: $H_b \ll H_a$.
A successful phenomenology thus results if the ratio
$H_b/H_a$ approaches a zero for a (small) range
of $b$ around the value $b_I$.  Let us expand the ratio
around this point
\be
{H_b \over H_a} = S + T\left(\frac{b}{b_I} - 1\right)^2 +...
\label{Hratio}
\ee
The dimensionless parameters $S$ and $T$ will be bounded
in size by the requirements that the spectral index $n$
of the density perturbations is close to 1, and that
sufficient e-folds of inflation occur to solve the horizon
and flatness problems.  Furthermore $\dot H_b$ must be
small as compared to $H_a^2$, otherwise the spectral index
would again too quickly deviate from 1.

To quantify these restrictions
let us return to the equations of motion for $a(t)$ and $b(t)$
and the general expression for the density perturbations.
Successful inflation requires that the number of efolds $N_e$
of superluminal expansion be sufficiently large.
If the initial and final values of the scale factors just before
and just after the inflationary epoch are denoted by subscripts
$(i,f)$ respectively, then
\bea
N_e & = & \int_{a_i}^{a_f} {da\over a}\\ \nonumber
& = &\int_{b_i}^{b_f} {H_a\over H_b} {db\over b}\\ \nonumber
& \simeq & \int_{b_i}^{b_f} {1\over S + T(b/b_I -1)^2} {db\over b}.
\label{Nefolds}
\eea
To a good approximation inflation begins and ends when the
condition $H_b/H_a\ll 1$ is violated.
Hence to obtain inflation, there are two possibilities.
Either $S$ must satisfy $S\ll 1$, which in
turn implies
\be
{b_{i,f}\over b_I} = 1 \mp T^{-1/2}.
\label{inflends}
\ee
The alternative is to have $S$ small (but not excessively)
while $T$ must be extremely small, which corresponds to slow,
power law, inflation.
Substituting these endpoint values into (\ref{Nefolds}) and performing
the integral gives the constraint on $S$ and $T$ arising from
$N_e$:
\be
N_e \simeq {1\over S+T}\left( {2\sqrt{T}\tan^{-1}(1/\sqrt{S})\over
\sqrt{S}} - \log(1+1/S) + 2 \log(1+1/\sqrt{T}) \right)
\label{efoldcond}
\ee
Requiring, say, $N_e>100$ then puts an upper bound on the
size of $(S,T)$.  In various limits this relation becomes
easy to state explicitly.  For example if $S=T\ll 1$ then
\be
N_e \simeq {\pi \over 2 T}
\ee
while if $S$ is very small, while $T$ remains $O(1)$,
\be
N_e \simeq {\pi \over \sqrt{S}} - \log(1/S).
\ee
In any case obtaining $N_e \sim 100$ efolds of inflation requires
either $S\simeq T \ltap 0.02$, or $T\simeq O(1)$, and $S\ltap 10^{-3}$.
So in practice to get a sufficient number of efolds $S$ should
be somewhere in the range $0.001 \ltap S \ltap 0.02$ (or, of course less)
depending on $T$.

Now let us turn to the constraint arising from the magnitude and
spectral index of the density perturbations.
To do this in detail, we must
make some assumptions about the identity of the inflaton.
Since we have argued that inflation should occur far away from the
eventual stabilization value of $b$, the most natural
candidate for the potential energy that drives inflation is
the radion potential itself, with therefore the radion
playing the role of the inflaton.
The correctly normalized field $\phi$ corresponding
to the scale factor $b$ is given by (see Appendix~A)
\be
\dot \phi = \sqrt{2n(n-1)} \mst^{(n+2)/2} b^{(n-2)/2} \dot b .
\label{bcanonical}
\ee
Substituting this into the expression for the density
perturbations (\ref{drho}) we
find
\bea
{\de\rho\over \rho} & = &  \frac{5}{12\pi\sqrt{2n(n-1)}} { H_a^2 \over
 \mst^{(n+2)/2} b^{(n-2)/2} \dot b} \\ \nonumber
& \simeq & \frac{5}{12\pi\sqrt{2n(n-1)}} { H_a \over
\mst (\mst b_I)^{n/2} S },
\label{drho2}
\eea
where in the last step we have used (\ref{Hratio}) and the fact
that we are close to $b_I$.
Requiring $b(t)$ to be essentially static during inflation,
and in particular that $H_b/H_a$ and $\dot H_b/H_a^2$ are small
near $b_I$ is equivalent, using the equations of motion, to
the statement that
\be
{2 \left( \frac{b}{n} \partial_b V - 2 V\right)\over 3
\left( b \partial_b V - (n-2) V\right)}\Biggr|_{b\simeq b_I} \ll 1.
\label{rhsconditions}
\ee
which translates into the condition that
\be
\left( \frac{b}{n} \partial_b V - 2 V\right)\Biggr|_{b_I} \ll V(b_I),\quad
b_I \partial_b V(b_I).
\ee
Therefore $(b \partial_b V)|_{b_I} \simeq 2n V|_{b_I}$, and
up to small corrections, the Hubble parameter $H_a$ during inflation
is given by
\be
H_a^2 \simeq {V(b_I)\over 6 b_I^n \mst^{(n+2)}}.
\label{Haeqn}
\ee
Finally this together with (\ref{drho2}) gives the expression
for the inflation generated density perturbations
\be
{\de\rho\over \rho} \simeq \frac{5}{12\pi} {1\over S (\mst b_I)^n}
\left({ V(b_I) \over 12n(n-1) \mst^4 }\right)^{1/2}.
\label{drho3}
\ee

The spectral index $n_\rho$ is defined by the comoving wavenumber
dependence
\be
{\de\rho\over \rho} \sim k^{(n_\rho -1)/2}
\label{nrhodef}
\ee
where at horizon crossing we have the relation $k/a = H_a$.  From
(\ref{nrhodef}) we can extract a more convenient expression for
$n_\rho$:
\bea
n_\rho -1 & = & 2 {d \log(\de\rho/\rho) \over d \log(a)} \\ \nonumber
& \simeq & 2 S {d \log(\de\rho/\rho) \over d \log(b)},
\label{nrhodef2}
\eea
where in the second line the parameterization for
$H_b/H_a$ (\ref{Hratio}) has been used.  Applying
this formula to the expression for the density perturbations
(\ref{drho3}) leads to
\be
n_\rho -1 = 2S\left( -n + {1\over 2 V}{d V\over d \log(b)}\right)
\label{nrhotheory} .
\ee
But from the two expressions for the ratio $H_b/H_a$, one
from the equations of motion (\ref{aeom},\ref{beom}), and the other
given by the
parameterization (\ref{Hratio}) we have, after some algebra,
\be
{1\over 2 V}{d V\over d \log(b)} = n - {n(n+2)\over 4}\left( S +
T (b/b_I -1)^2 \right)
\ee
and therefore
\be
n_\rho -1 \simeq -{n(n+2)\over 2}\left( S^2 +
ST (b/b_I -1)^2 \right)
\label{nrhofinal}
\ee
This expression together with the experimental constraint
$|n_\rho-1| < 0.2$ forces $S\ltap 0.1$, thus ruling
out the case of $T\ll 1$, $S\sim O(1)$ allowed by the
earlier condition of a sufficient number of efolds
of inflation.  Note
however that the solutions with $T\sim O(1)$ and $S\ltap 0.01$
automatically satisfy $|n_\rho-1| < 0.2$ over essentially
the whole range of $a(t)$ inflation.  (Experimentally
all we require is that $|n_\rho-1| < 0.2$ in the range
of scales between the COBE and large-scale structure
measurements -- roughly 10 efolds rather than the full
duration of inflation.)

Finally we can use the magnitude of the measured CMB fluctuations
to constrain the size $b_I$ of the internal dimensions during the
inflationary epoch.
Recall that COBE and the other CMB measurements tell us that
at the time the scales $k\simeq 7 H_{\rm now}$ were being inflated
outside of the horizon, the density perturbations were of size
\be
{\de\rho\over \rho}\Biggr|_{\rm COBE} \simeq 2 \times 10^{-5}.
\label{COBE}
\ee
This together with the formula for $\de\rho/\rho$ (\ref{drho3}),
and the bound on the parameter $S$ leads to a constraint on a
combination of the size of the potential during inflation
$V(b_I)/\mst^4$ and the volume $V_I = (b_I)^n$
of the internal dimensions, both in units of
the fundamental Planck scale:
\be
V_I \mst^n \simeq 5\times 10^4 \left({0.02\over S}\right)
\left({V(b_I)\over \mst^4}\right)^{1/2}.
\label{inflsize}
\ee
For $S\simeq T\simeq 0.02$ and $V(b_I)$ having a perfectly
reasonable value of $V(b_I)\simeq (200\gev)^4$, (for example
if $\mst\simeq 1\tev$) we therefore
discover that the epoch of inflation generating
COBE needs to occur around the value
$b_I \simeq 10^{3/n} \mst^{-1}$,
\ie, when the internal dimensions are still relatively small
as expected.

Now let us return to the justification of the use of the
usual 4-dimensional expression (\ref{drho}) for the density
perturbations.  If the de Sitter horizon $(H_a)^{-1}$
during inflation
is much smaller than the size of the internal dimensions then
the full $(4+n)$-dimensional expression for the density
fluctuations must be employed, while if $(H_a)^{-1} \gg b_I$
then it is correct to use the long-distance, effectively
4-dimensional, expression.   From the constraint arising
from $\de\rho/\rho$ (\ref{inflsize}), and the expression for
$H_a$ during inflation (\ref{Haeqn}) we find that
\be
H_a b_I \simeq {(5\times 10^{-3})^{(n-2)/n} \over \sqrt{6}}
\left({S\over 0.02}\right)^{(n-2)/2n}
\left({ V(b_I)\over \mst^4}\right)^{(n+2)/4n}.
\label{HabI}
\ee
Thus $H_a b_I$ is always substantially less than 1 for the
range of parameters of interest, and the 4-dimensional
description of the generation of $\de\rho/\rho$ is
correct.

\subsection{Inflation from the Einstein frame perspective}

Most dynamical aspects the world-as-a-brane scenario
are the most transparent in terms of the geometric variables
employed so far, which can be referred to as the string-frame
quantities. The reason is that the kinematics in this frame
is automatically expressed in terms of the units felt by the
observers which live on the wall. In particular, such observers
choose to define scales using their own masses and Compton wavelengths
as yard sticks, and in the string frame these yard sticks are time
independent. However, it is illustrative to consider geometrical
evolution of the universe in light of the reference
frame where the gravitational sector of the theory coincides
with classical Einstein gravity. In this frame, the equations
of motion resemble the coupled gravity-matter models
considered in the context of non-minimal theories of
gravity \cite{bdinf},
and the results obtained from them can be a useful addition
to physical intuition.

For this reason, here we will recast the description of the inflationary
stage into the Einstein conformal frame of the theory
on the brane (\ref{bulkS}), (\ref{actions}). This frame
is defined by the requirement that the
propagator of the graviton is canonical. The map which
casts the kinetic term of the graviton into the canonical
form is
\be
\bar g_{\mu\nu} = \left(\frac{\mst}{\mpl}\right)^2 (\mst b)^n g_{\mu\nu}
\ee
Explicitly, we can define the Einstein frame comoving time
and scale factor according to
\be
d\bar t = \frac{\mst}{\mpl} (\mst b)^{n/2} dt ~~~~~~~~~~~~
\bar a = \frac{\mst}{\mpl} (\mst b)^{n/2} a
\ee
where all barred quantities refer to the Einstein frame. Now, the radion
field is equivalent to a scalar field, defined by
\be
\mst^{n+2} b^n = \mpl^2 \exp\left(-\sqrt{\frac{n}{2(n-1)}} \frac{\phi}{\mpl}
\right)
\ee
Note that the normalization employed here is
dictated by the definition of the perturbation
of the radion away from its mean value during
inflation, which sets the effective
background Planck mass.
We note that the initial conditions for inflation in
the Einstein frame, in the units of the effective Planck scale,
require homogeneity and flatness
over distances of $\sim l_{Pl,eff} =1/\mpl$, which can be
recognized as a usual inflationary initial condition.

This Einstein frame picture is very useful to compute the
density contrast in models where radion is the inflaton.
The density contrast is given by the standard formula
\be
\frac{\delta \rho}{\rho} = \frac{\bar H^2}{2\pi \phi'}
= 8 \frac{\bar H^2}{\phi'}
\label{econtr}
\ee
in our normalizations, and
where the prime is the derivative with respect to $\bar t$.
To determine the density contrast in the string frame, we
recall that it is, roughly, conformally invariant
during inflation \cite{kko}, and
conformally transform (\ref{econtr}) to the string frame.
The radion and the Hubble parameter transform to
\bea
\phi'&=& - \sqrt{2n(n-1)} \frac{\mpl^2 H_b}{\mst(\mst b)^{n/2}}
\nonumber \\
\bar H &=&
 \frac{\mpl}{\mst(\mst b)^{n/2}}
\left(H_a + \frac{n}{2} H_b \right)
\eea
which is straightforward to determine from their
definitions. With this, we find
\be
\frac{\delta \rho}{\rho} = \frac{8}{\sqrt{2n(n-1)}}
\frac{(H_a + \frac{n}{2}H_b)^2}{\mst(\mst b)^{n/2} H_b}
\ee
Now we consider the tilt of the perturbation spectrum.
To consider it, we define the effective slope by
\be
\theta \sim \frac{\frac{\delta \rho}{\rho}(t_a)}{
\frac{\delta \rho}{\rho}(t_b)} \sim \left(\frac{b(t_b)}{b(t_a)}\right)^{n/2}
\ee
where the last equality arises
because of the slow roll conditions $H_{a,b}(t_a) \sim H_{a,b}(t_b)$ during
inflation. Since $b(t_a) \sim b(t_b)(1+H_b \delta t)$ and since we compare
the tilt between $50$ and $60$ efoldings, we get $\delta t \sim 10 H^{-1}_a$.
Thus roughly, $\theta \sim 1- 5n \frac{H_b}{H_a}$
On the other hand, during this time $\theta$ changes according to
$\theta \sim \left(\frac{k_a}{k_b}\right)^{(n_\rho-1)/2}$,
and given that $k_k \sim a_k H_a$, we find that the ratio of the
wave vectors is given by
$k_a/k_b \sim a_a/a_b \sim \exp(H_a \delta t) \sim \exp(10)$.
Taking this and the bounds on the spectral index $n_\rho$,
$n_\rho \le 1 \pm 0.2$, we find that
\be
\theta \sim e^{\pm 1/2}.
\ee
This leads to the following inequality which the
expansion rates must satisfy during
inflation:
\be
H_a \gtap \frac{5n}{1-\exp(\pm 1/2)} H_b \ge 15 n H_b ,
\ee
As a result, the density contrast $\delta\rho/\rho$ is
\be
\frac{\delta \rho}{\rho} \simeq \frac{8}{\sqrt{2n(n-1)}}
\frac{H^2_a}{\mst(\mst b)^{n/2} H_b}
\ee
The COBE data tell us that $\delta \rho/\rho \sim 10^{-5}$.
Using this and $H_a \ge 15n H_b$, we find
\be
H_a \ltap 10^{-7} \mst (\mst b)^{n/2}
\ee

We can use this inequality to obtain a bound on the size
of the internal dimensions $b_I$ during inflation.
We repeat that the evolution is similar to the
Einstein frame thanks to the slow roll conditions.
Hence the vacuum energy density
during inflation is
\be
\rho_{\rm vacuum} \sim \mst^{(2+n)} b_I^n H^2_a
\ee
Using our estimate for $H_a$,
\be
\rho_{\rm vacuum} \sim 10^{-14} \mst^4 (\mst b_I)^{2n}
\ee
However $\rho_{\rm vacuum}$ must not exceed $\mst^4$.
In fact, since initially the theory
is very close to the quantum gravity scale, in order to insure the validity
of the semiclassical approximation the energy density cannot exceed,
let's say, $10^{-5} \mst^4$, meaning that the energy scale is about a
factor of $10^{5/4}$ below the quantum gravity scale.  This ensures
that the semiclassical description is correct. Using the equation
for the energy density in terms of $b_I$ above, we obtain the inequality
\be
b_I \ltap 10^{9/2n} \mst^{-1}
\label{binit}
\ee
Note that the upper limit on $b_I$ ranges between about $200 \mst^{-1}$
for $n=2$ to about $6 \mst^{-1}$ for $n=6$. In either case however,
these are clearly the correct initial conditions for inflation, which
in fact come naturally in this context.

\section{Post-inflation evolution to stabilization point}

As we have discussed in the previous section, when the
inflationary stage ends, the ratio of internal space
to on-brane Hubble rates of expansion $H_b/H_a$
approach unity.  From this time on, the slow roll conditions
for the effective potential cannot be upheld any more.
The kinetic energy stored in the
expansion of the internal space begins to play a significant role
in the evolution of the brane-world. Rather interestingly,
however, we will see that under rather general conditions
the expansion of the whole
higher-dimensional universe under
the combined influence of the potential and kinetic energies
during this era is well approximated by
a generalization of the
well-known Kasner solutions. These
solutions generically describe
cosmic evolution which is anisotropic in different directions,
with a subset of the directions {\it contracting} while the
others expand.  Specifically we will find that after the end
of the initial brane-world inflation the
directions longitudinal to the brane contract,
while the internal directions expand, both
according to some power-law time dependence. The precise form of
the power laws is controlled by the codimension
of the brane ($n$ in our notation) and the leading-order
behavior of the stabilizing potential $V(b)$ as a function
of the scale factor $b$ of the internal dimensions.

\subsection{Theory of the era of contraction}

To illustrate generic features of such behavior,
and show why it should be immediately expected, at least
in a subset of cases, consider the limiting case
where after the exit from the inflationary
stage the potential $V(b)$ drops by many orders of magnitude.
The evolution of the scale factors $a(t)$ and $b(t)$ is
then controlled entirely by the kinetic energy, and the
equations of
motion appropriate for this case are just the
higher-dimensional vacuum Einstein equations,
\be
R_{\mu\nu} = 0
\label{eeqs}
\ee
In other words (\ref{aeom},\ref{beom}) with $V$ set to zero.
The $(4+n)$-dimensional solutions which are
consistent with the brane symmetries are of the form
\be
ds^2 = - dt^2 + a^2_i \Biggl(\frac{t}{t_i}\Biggr)^{2 k} d\vec x^2_3
+ b^2_i \Biggl(\frac{t}{t_i}\Biggr)^{2 \ell} d\vec y^2_n
\label{kasner}
\ee
where $a_i, b_i$ are now the ``initial'' values of the scale factors
as set by the end of the inflationary stage.  The powers $k$ and $\ell$
are uniquely determined by the Einstein equations, which reduce to
two simple algebraic equations,
\bea
3k + n\ell &=& 1 \nonumber \\
3k^2+n\ell^2 &=& 1
\label{kaspower}
\eea
A metric of the form (\ref{kasner}) with exponents satisfying
(\ref{kaspower}) is known as the Kasner solution.
The solutions of the algebraic equations for the exponents are
\bea
&&k = \frac{3 \mp \sqrt{3n(n + 2)}}{3(n+3)} \nonumber \\
&&\ell = \frac{n \pm \sqrt{3n(n + 2)}}{n(n + 3)}
\label{kasnercoeff}
\eea
Phenomenologically, we certainly need the internal dimensions
to grow in size and approach the stabilizing value, and this
selects the upper sign in the equalities (\ref{kasnercoeff}),
which then implies that our longitudinal brane directions contract.
As we will soon see when we re-introduce the potential $V(b)$, this
behavior is physically selected by the asymmetry embodied in the
potential, and in particular the fact that it must have a
minimum at the stabilizing value $b_0$, with
$b_0^n = \mpl^2/\mst^{n+2}$.
In any case in this simple potential-free Kasner case,
the explicit values for the powers range between
$k = -0.1266, \ell= 0.69$ for $n=2$, and $k = -1/3, \ell = 1/3$ for $n=6$.

It is useful to express the brane scale factor
$a(t)$ as a function of the size of the internal
dimensions $b(t)$,
\be
\frac{a}{a_i} = \left(\frac{b}{b_i}\right)^{-|k|/\ell}
\label{abeq}
\ee
The amount of contraction of our scale
factor is thus determined by the increase in $b$
(which in turn is determined once the end of inflation
value of $b=b_i$ is specified), and the known power
$-|k|/\ell$.
Note that for $n=2$, $k/\ell \sim -0.183$, and so
the brane world contracts by only one order of magnitude for
almost six orders of magnitude of radion increase.  As $n$ increases,
this dependence speeds up, and for $n=6$ the contraction of the
brane dimensions and expansion of the internal dimensions are
essentially equal in magnitude (we have of course ignored small
transient effects at both the start and end of this Kasner phase which
slightly modify the above relationships).

We are typically interested, however, in the case
where there exists a non-trivial $V(b)$ potential
as well as matter on our brane.
We again take the metric to be of the form
\be
ds^2 = -dt^2 + a^2(t) d\vec x^2_3 + b^2(t) d\vec y^2_n
\label{metric}
\ee
and assume that the matter on the wall is
represented by a classically conserved, perfect
fluid energy-momentum tensor $T_{\mu\nu} = (\rho +p)
u_\mu u_\nu + p g_{\mu\nu}$, $\nabla_\mu T^{\mu\nu} = 0$,
where $u_\mu$ is a future-oriented timelike vector,
with the components
$u^\mu = (1, \vec 0_{3+n})$ in the basis (\ref{metric}).
Here $\rho$ is the energy density of the {\it wall} matter and
$p = \gamma \rho$ is the pressure, with $\gamma$ a constant
given by the speed of sound on the wall.  The equations of motion
slightly generalize (\ref{aeom}) and (\ref{beom}),
\bea
&&6 H_a^2 + n(n-1) H_b^2 + 6n H_a H_b = \frac{V + \rho}{\mst^{n+2} b^n}
\nonumber \\
&&\frac{\ddot b}{b} + (n-1) H_b^2 + 3 H_b H_a
= \frac{1}{\mst^{n+2} b^n} \left(\frac{2 V}{n+2}
- \frac{b}{n(n+2)} \frac{\partial V}{\partial b} +
\frac{\rho-3p}{2(n+2)} \right) \nonumber \\
&& \frac{\ddot a}{a} + 2H_a^2 + n H_b H_a = \frac{1}{\mst^{n+2} b^n} \left(
\frac{b}{2(n+2)} \frac{\partial V}{\partial b}
- \frac{n-2}{2(n+2)} V +
\frac{\rho + (n-1) p}{2(n+2)} \right) \nonumber  \\
&&\dot \rho + 3H_a(p + \rho) = 0.
\label{eoms}
\eea
The last equation is of course the usual wall
energy-momentum conservation equation,
$\nabla_\mu T^{\mu\nu} = 0$.

It is important to note that the contraction of the brane
in Kasner-like solutions
leads to an increase of the energy density of any matter
or radiation present on the brane at the beginning of the
contraction, and this will lead to ``bounce'' solutions
that will be important to us later.
Also note that these equations generically
receive quantum corrections from particle
production via curved space effects or the conformal anomaly.
The corrections would manifest themselves as
a nonzero source term on the RHS, which as we will see would
lead to significant effects only early on, and therefore could
be modeled by an appropriate choice of the initial condition for $\rho$.

Namely, the energy density on the wall corresponds to the particles
produced by changing gravitational fields at the
end of inflation. Since these phenomena are essentially similar to
Hawking radiation, the value of $\rho$ as compared to the potential
is suppressed by a factor of $H_a^2/M^{n+2} b^n$ at the end
of inflation. Using the COBE data to constrain this quantity, we
see that initially $\rho$ is much smaller than $V$. Hence ignoring
it is an excellent approximation. For most of this stage, therefore,
the dynamics of the universe is determined by the interplay of the
radion kinetic and potential energy, which completely determine the
evolution of the wall geometry. We will simply carry on
with the analysis of (\ref{eoms}), treating $\rho$ as a small
perturbation and ignoring its back-reaction on $a$ and $b$.
However, although to the lowest order, the particle production
phenomena are negligible, they could lead
to interesting effects for reheating.
We will return to this later.

However before we consider such blue-shifting of the brane-localized
energy density, let us turn our attention to the more general and
appropriate case of non-negligible radion potential.  Remarkably
in this case a form of Kasner-like behavior still applies.

Quite generically,
in the semiclassical limit, fully valid at this stage of the evolution,
the potential may be viewed as an expansion in inverse powers
of the radion field $b$.  At generic values of $b$ away from the
stabilization point the
potential will be dominated by a single term in this expansion.
Hence, we can simply approximate $V(b)$ by a monomial of the
form $V = W b^{-p}$, where $W$ is a dimensionful parameter with
$[W] = 4-p$.\footnote{$W$ could well depend
logarithmically on $b$.  This mild additional $b$
dependence will not change either our qualitative, or to a good
approximation our quantitative conclusions.  For simplicity we
ignore it in the following discussion.}

If we substitute $V = W b^{-p}$ in the equations of motion, ignoring
$\rho$ and $p$ for the moment we find
\bea
&&6 H_a^2 + n(n-1) H_b^2 + 6n H_a H_b =
\frac{W}{\mst^{n+2} b^{n+p}} \nonumber \\
&&\frac{\ddot b}{b} + (n-1) H_b^2 + 3 H_b H_a=
\frac{(2n+p) W}{n(n+2) \mst^{n+2} b^{n+p}}
\nonumber \\
&& \frac{\ddot a}{a} + 2H_a^2 + n H_b H_a =
- \frac{(n+p-2)W}{2(n+2)\mst^{n+2}b^{n+p}}.
\label{eomscoast}
\eea
These equations can in fact be solved exactly!  With appropriate
field redefinitions and gauge (coordinate) transformations, they
can be mapped to a system of equations describing the motion of two particles
in one dimension, one free, and another Liouville with an exponential
potential.

First note that in a certain parameter range these exact solutions
asymptotically converge to the ``potential-free'' Kasner solutions
(\ref{kasner}) and (\ref{kasnercoeff}) above, as we
discuss in Appendix~B.  This parameter range
turns out to be the one in which, upon substitution of the power dependence
(\ref{kasnercoeff}) into the equations (\ref{eomscoast}), the potential
terms on the RHS vanish more quickly as a function of time than the
LHS.  Since the LHS always scales as $\sim t^{-2}$ this is the case when
$(n+p)\ell > 2$, or equivalently
\be
{(n+p)\over n(n+3)} ( n+ \sqrt{3n(n+2)} >2
\label{simplekasner}
\ee
This gives a curve in $(n,p)$ space above which (\ie, for larger values
of $p$) the exact solution asymptotes to the potential-free Kasner
exponents.  The critical values of $p$ vary from $0.899$ at $n=2$ to
$p=0$ at $n=6$.

We now discuss the changes of variables which allow the exact solution
of the equations with potential.  First define
\be
a = a_i e^{\alpha(t)} \qquad b = b_i e^{\beta(t)}
\label{redefs}
\ee
where $a_i$ and $b_i$ are as before the initial values of the scale factors
at the beginning of the epoch of radion coasting.  This means that
the appropriate initial conditions for the dimensionless variables
$\alpha$ and $\beta$ are $\alpha_i = \beta_i = 0$.  Further,
defining the parameter $\omega = \frac{W}{\mst^{n+2} b^{n+p}_i}$,
substituting this and (\ref{redefs}) into (\ref{eomscoast}),
and going to the new time variable $\tau$:
\be
d\tau = -e^{-3\alpha - n \beta} dt
\label{newtime}
\ee
leads after some simple algebra to the equations of motion
in their final form, suitable for explicit analysis:
\bea
&&6 \alpha'^2 + n(n-1) \beta'^2 + 6n \alpha' \beta'
= \omega e^{6\alpha + (n-p) \beta} \nonumber \\
&&\beta'' = \frac{(2n+p) \omega}{n(n+2)} e^{6\alpha + (n-p)\beta}
\nonumber \\
&& \alpha'' = - \frac{(n+p-2)\omega}{2(n+2)} e^{6\alpha +(n-p) \beta} .
\label{dimeomsfin}
\eea
Here primes denote derivatives with respect to $\tau$.
The specific form of $\tau = \tau(t)$ can be determined after the
solutions are found.

The equations (\ref{dimeomsfin})
are immediately integrable.  Indeed, consider the linear combination
$4\alpha + \frac{2n(n+p-2)}{2n+p} \beta$.
By using the second order differential
equations, it is easy to verify that
$4\alpha'' + \frac{2n(n+p-2)}{2n+p} \beta''=0$.
Hence we can immediately write one first integral of (\ref{dimeomsfin}):
\be
4\alpha + \frac{2n(n+p-2)}{2n+p} \beta = C_1 + C_2 \tau
\label{oneintegr}
\ee
where $C_1$ and $C_2$ are integration constants to be determined
later.

To find the other integral of motion, we can define the new
variable
\be
X = 6 \alpha + (n-p) \beta
\label{defx}
\ee
The remaining independent second order
differential equation becomes
\be
X'' = \frac{6n-4np-n^2 - p^2}{n(n+2)} \omega e^X
\label{liouville}
\ee
This is the Liouville equation, corresponding to a particle
in 1 dimension moving in an exponential potential.  This has
an easily determined first integral,
simply given by the conservation of energy:
\be
X'^2 = 2 \frac{\Delta}{n(n+2)} \omega e^X + E_0
\label{liouenergy}
\ee
with $E_0$ the energy integral, and where we have
defined $\Delta = 6n-4np-n^2-p^2$.
At this moment, we need to make three observations. First,
$X$ is a good independent variable only as long as
$\Delta$ is nonzero. If it is zero, the variable $X$
degenerates to the previous first integral (\ref{oneintegr}), up
to an overall constant, and so another independent integral
should be used.  This is easy to take into account however, and
besides corresponds to a set of measure zero in the phase space
of solutions, and thus we will not pay it much attention here.
We will instead focus on the more generic cases where $X$ is
independent from (\ref{oneintegr}).

Second, the curve in the $(n,p)$ plane where $\De=0$ vanishes is
precisely the curve defined by (\ref{simplekasner}) that separates
the traditional potential-free Kasner solutions from the more
general (but still asymptotic power-law) behavior that we discuss below.
In particular, the traditional potential-free Kasner solutions
apply asymptotically in the region $\De < 0$.  The full behavior
in this region, including the transient regime before the power-law
dependence on $t$ sets in can easily be discussed by a simple
generalization of the analysis described below and in Appendix~B
for the case $\De>0$.  Leaving the region $\De<0$ to the Appendix,
we focus on the novel case of $\De >0$ in the following.

Third, the integrals of motion $C_2$ and $E_0$ are not
independent by virtue of the Einstein
constraint equation, which is the first equation of
(\ref{dimeomsfin}). If we take (\ref{liouenergy})
and the first derivative of (\ref{oneintegr}), and substitute
them into the constraint, we find
\be
E_0 = \frac{3(2n+p)^2}{4n(n+2)} C^2_2.
\ee
Thus we see that the constant $C_2$ completely
controls the dynamics. With this, we have essentially
reduced the system (\ref{dimeomsfin}) to a functional
constraint (\ref{oneintegr}) and a simple $1^{\rm st}$ order equation
\be
X'^2 = 2 \frac{\Delta}{n(n+2)} \omega e^X
+ \frac{3(2n+p)^2}{4n(n+2)} C^2_2
\label{lasteq}
\ee
which can be easily integrated when
$\Delta \ne 0$.\footnote{If $\Delta = 0$ a simple modification
of this procedure still yields the second integral of motion.}

As shown in Appendix~B when $\Delta >0$, the exact solutions of
these equations yield the long-time behavior of the scale
factors $a(t)$ and $b(t)$.  It is given by
\bea
a &=& a_i \left(\frac{t}{t_i}\right)^{-\frac{n(n+p-2)}{(n+p)(2n+p)}}
\nonumber \\
b &=& b_i \left(\frac{t}{t_i} \right)^{\frac{2}{n+p}} .
\label{solsonefin}
\eea
These solutions describe a situation in which the on-brane
scale factor shrinks while the size of the internal
dimensions continues to grow.  The exact solutions
show that for $C_2=0$ they are generally valid, while
they are asymptotic long-time attractors for the generic cases
with $C_2<0$ (\ie, the case in which both $H_a$ and $H_b>0$ after
the end of inflation).  This means
that even if the brane scales initially continue to expand, soon
after the end of inflation they inevitably start to contract.
Note that the $t$-dependence of $b(t)$ is exactly such as
to have the potential-dependent RHS of the evolution equations
scale in the  same $1/t^2$ fashion as the LHS.

In any case, the main point of this analysis is that the
evolution after the initial stage of inflation continues
into a phase of slow progress of $b(t)$ towards the
stabilization point, with, under quite general conditions,
{\it a simultaneous contraction of our brane scale factor} $a(t)$.

The amount of $a(t)$ contraction
is controlled by the exponents in (\ref{solsonefin})
(or in the case of $\Delta<0$ the exponents in (\ref{kasnercoeff})),
and the amount of expansion of $b$ until the stabilization point
is reached.
We can use the asymptotic form of the solutions
to place an upper bound on the amount of contraction of the brane
as a function of the evolution of $b$.  We have
\be
\frac{a}{a_i} \le \left(\frac{b_i}{b}\right)^\zeta
\label{contrrad}
\ee
where the parameter $\zeta$ is given by
\bea
\zeta &=& \frac{n(n+p-2)}{2(2n+p)} \qquad {\rm for} ~\Delta > 0
\nonumber \\
\zeta &=& \frac{3n-\sqrt{3n(n+2)}}{6} \qquad
{\rm for} ~\Delta < 0.
\label{contrrad2}
\eea
In these equations recall that $\Delta = 6n - 4np - n^2 - p^2$,
while the effective 4-dimensional potential for $b$ varies
as $V\sim b^{-p}$.  It is easy to see that the exponents for the
$a(t)$ and $b(t)$ evolution calculated in the $\De>0$ and $\De<0$ regions
are continuous on the curve $\Delta =0$, and thus so is $\zeta$.
Furthermore, note that for a given $n$ the greatest $a(t)$ contraction
occurs in the $\Delta<0$ case.

\subsection{Phenomenology of the era of contraction}

There are a number of interesting consequences of the
period of $a(t)$ contraction just described.
First, during the contraction era, the brane
universe looses a number of efoldings of the scale factor to the shrinkage.
This implies that the early period of inflation needs to produce
that many efoldings more than the naive minimum needed to solve the
horizon and flatness problems.
Moreover, the contraction of $a(t)$ blue-shifts any energy
density left on our wall at the end of inflation, and as
we will argue below, this could, together with the remnant of
the de Sitter era Hawking radiation, conceivably be the source
of reheating of our brane.

To quantify these remarks, we note that the contraction factor of the brane
cannot exceed the amount
\be
\frac{a_f}{a_i} \le \left(\frac{b_i}{b_0}\right)^\zeta
\label{shrinkage}
\ee
where the fact that the final value of $b(t)$ is well-approximated
by the stabilizing value, $b_0$, has been used.  The value of $b_i$
is set by the size of the internal dimensions at the end of the
early period of inflation and at worst is of order the fundamental
Planck length $b_i\ge \mst^{-1}$.  (The COBE constraint
(\ref{inflsize}) typically
requires a larger value of $b_i$, and this just lessens the amount
of contraction.)  Thus, the maximum amount of contraction is bounded
from above by
\be
\frac{a_f}{a_i} \le \left(\frac{\mst}{\mpl}\right)^{2\zeta/n} .
\label{shrinkagebound}
\ee
The numerical value of this formula ranges between about
$900$ for $n=2$ to about $2\times 10^5$ for $n=6$.  This means that we
loose to contraction at most about $7$ efoldings of inflation
for $n=2$ up to about $12$ for $n=6$.  Clearly, this is not excessive,
and can be easily made up for by a slightly longer period of early
inflation.

We now turn to the evolution of the wall and bulk energy densities
during the period of contraction.
First note that the dominant form of energy density at the end
of the inflationary period is the bulk energy of the radion,
\ie the $b(t)$ scale factor kinetic and potential energy.
The (effective 4-dimensional) kinetic energy density of
the radion is
\be
\rho_{b,KE} = n(n-1)\mst^{n+2} b^n (H_b)^2
\ee
while its 4-dimensional potential energy density is just $V(b)$
as used in the previous sections.
>From the exact solutions presented in the previous
subsection we find that scaling of the energy
density of the radion depends on the exponents $(k,\ell)$ for
the evolution of the scale factors $a\sim t^k$, and $b\sim t^\ell$,
which in turn depend on whether the potential is important or not,
in other words the sign of $\Delta$.  The result is
\bea
&\Delta \ge 0& ({\rm potential~case}) \\ \nonumber
&&\rho_{b,PE}\sim b^{-p},\quad \rho_{b,KE}\sim b^{-p} \\ \nonumber
&\Delta <0& ({\rm simple~Kasner})  \\ \nonumber
&&\rho_{b,PE}\sim 0,\quad \rho_{b,KE}\sim
b^{2n-\sqrt{3n(n+2)}} .
\label{radionenergy}
\eea

So far we haven't discussed what ends the contraction period.
As shown in detail in Appendix~C, it is a remarkable
fact that contraction
of $a(t)$ stops and reverses when $\rho_{wall}$
satisfies\footnote{The exact analysis we have
performed proves that this is the
case when the dominant form of radion energy is kinetic,
and, moreover, as we will show in Section~3.4, necessarily
and automatically occurs if the radion stabilizes.}
\be
\rho_b = n(n-1) \mst^{n+2} b^n H^2_b = \rho_{wall}.
\ee

There are two generic possibilities for how this
condition may come to be satisfied:  The first takes the
primordial $\rho_{wall}$ left over from the inflationary
epoch.  If there is sufficient $a(t)$ contraction,
this inflationary $\rho_{wall}$ can becomes
comparable to $\rho_b$ before $b$ reaches the stabilization
point, $b_0$.  Then a ``Big Bounce'' occurs, the
contraction stops, and a modified expansionary phase that we
discuss below begins.  This expansion finally becomes the usual
FRW expansion after $b$ reaches $b_0$.
The second possibility is that a form of reheating takes place
on the wall which is totally unconnected with the contraction of $a(t)$,
but that again leads to $\rho_{wall} \ge \rho_b$.
Possibilities in this class include the decay of some metastable
state on the wall, or the collision of some other brane
with our brane.

Consider the first, model-independent, scenario.  We are used
to thinking of the universe at the end of inflation
as being very cold, but due to the de~Sitter era Hawking
radiation this is not completely so.
What is the initial value of the ratio $\rho_{wall}/\rho_b$ so
produced?  At the end of inflation, our brane is not
empty but inevitably contains a radiation bath left over from the
early inflationary de~Sitter phase, whose
energy density at the end of the de~Sitter phase
has been estimated by Ford \cite{ford} (see also \cite{parker}),
with the result
\be
\rho_{wall,i} \simeq 10^{-2} (H_{a,I})^4 ,
\label{initrhow}
\ee
depending on the exact nature (conformal or not) of the coupling
between the matter and metric.
Here $H_{a,I}$ is the Hubble parameter at the exit from inflation.
On the other hand, the initial value of the radion energy
is of order
\be
\rho_{b,i} \simeq n(n-1)\mst^{n+2} (b_I)^n (H_{b,I})^2
\simeq 6\mst^{n+2} (b_I)^n (H_{a,I})^2,
\label{initrhob}
\ee
since inflation terminates when $H_b/H_a$ approaches unity.
Furthermore due to the slow roll conditions, at inflation exit
$H_a$ is roughly the same as during the inflationary phase.
Given these expressions for the initial energy
densities, equations (\ref{Haeqn}) and (\ref{drho3}) together with
the constraint of reproducing the CMB/COBE data lead
to the interesting relationship
\be
{\rho_{wall}\over\rho_b}\Biggr|_i \simeq \frac{n(n-1)}{6} S^2
\left({\de\rho\over \rho}\Biggr|_{\rm COBE}\right)^2 .
\label{initrhoratio}
\ee

Due to the contraction
of our scale factor, the wall-localized radiation can become
important, the ratio $\rho_{wall}/\rho_b$ approaching unity.
Using the fact that $\rho_{wall} \sim a^{-4}$, and the relation
(\ref{contrrad}) between the evolution of $a$ and $b$, we find that
it scales as
\bea
&{\rho_{wall}\over\rho_b}\sim &
{b^p\over a^4}\sim b^{(2n^2+4np+p^2-4n)/(2n+p)}
\qquad {\rm for} ~\Delta > 0
\nonumber \\
&{\rho_{wall}\over\rho_b}\sim & {b^{-2n+\sqrt{3n(n+2)}}\over a^4}\sim
b^{\sqrt{n(2+n)/3}}
\qquad {\rm for} ~\Delta < 0.
\label{rhoscaling}
\eea

Given the initial value (\ref{initrhoratio}) of the ratio we
can compute with the aid of the scaling laws (\ref{rhoscaling})
how it behaves as the contraction proceeds.
Putting in the experimentally observed
value of $\de\rho/\rho\sim 2\times 10^{-5}$, and taking
a reasonable value of $S\sim 0.01$, we find that in the
simple Kasner $\Delta<0$ case, $\rho_{wall}=\rho_b$
before $b\simeq b_0$ for all $n=2,\ldots,6$.  On the other hand,
in the case of the Kasner-like solutions with potential,
$\Delta>0$, the rate of increase of the ratio
$\rho_{wall}/\rho_b$ is always slower as a function of $b$,
and in some cases (\eg, $n=3$, $p=0$) the stabilization point is
reached before $\rho_{wall}\sim\rho_b$.
We will focus on the $\Delta<0$ cases for illustration in the
following.

Specifically, using the scaling laws for the evolution of the energy
densities, $\rho_{wall}=\rho_b$ at a value of $b=b_f$ given by
\be
\left(\frac{b_f}{b_i}\right) = \left(
S \frac{\de\rho}{\rho}\Biggr|_{COBE}\right)^{-2\sqrt{n(n+2)/3}}.
\ee
We can now compute the energy density on the wall at the end of contraction:
\be
\rho_{wall,f} \simeq  7\times 10^2 \mst^4 (b_I \mst)^{2n} n(n-1)
\left( S \frac{\de\rho}{\rho}
\Biggr|_{COBE}\right)^{(2(n+1)^2 - 4n\sqrt{n(n+2)/3})}.
\label{reheatbound}
\ee
In terms of the temperature of radiation at the end of the contraction
phase, this becomes
\be
T_{wall,f} \simeq  6\mst (b_I \mst)^{n/2} (n(n-1))^{1/4}
\left( S \frac{\de\rho}{\rho}
\Biggr|_{COBE}\right)^{((n+1)^2/2 - n\sqrt{n(n+2)/3})}.
\label{tempbound}
\ee
If we use the fact that the typical volume (measured
in units of the fundamental Planck mass $\mst$) of
the internal dimensions during inflation is at most
$(b_I \mst)^{n}\simeq 10^3$, and that
$S\de\rho/\rho < 2\times10^{-7}$, given COBE and $S\ltap 0.01$,
we find that the wall temperature at the end of contraction
is bounded above by $1.4\times 10^{-6},7\times 10^{-7},
4.5\times 10^{-6}, 3\times 10^{-4}$, and $0.22$ in units
of $\mst$, for $n=2,...,6$ respectively.

We compare this to the limits found in \cite{ADDlong}, which
showed that the temperature of radiation on the brane, {\it by
stabilization}, cannot
exceed a certain maximal value in order to prevent overproduction
of bulk gravitons by evaporation.  This temperature, called the
normalcy temperature $T_*$ in \cite{ADDlong}, was defined as the
temperature above which the
cooling of the wall by graviton production begins to compete with the
normal adiabatic cooling by expansion.  It was estimated to
range between about $10\mev$ for $n=2$ to about $1\gev$ for
$n=6$.  For $\mst\sim 1\tev$,
the above expression for the reheating temperature of the wall radiation
at the end of contraction gives about $1.4\mev$ for
$n=2$ up to about $300\mev$ for $n=5$ and $220\gev$
for $n=6$.  In fact, if we take $\mst\simeq 30\tev$ for the $n=2$ case,
as suggested in \cite{ADDlong}, we can see
that the upper bound on the predicted wall reheating temperature
becomes $\simeq 45\mev$. Thus we see that apart from the $n=6$ case
we certainly satisfy the normalcy bounds on the reheating temperature.

However, these numbers are only the upper bounds for a number of
reasons.  First, in reality, the evolution is not always Kasner-like.
Instead, there are periods immediately
after the end of inflation when the brane scale factor is increasing,
or decreasing more slowly than in the Kasner case.
As a result of this, the actual blue-shift is less than the bounds
used above, and this can easily lower the upper bound by an $O(1)$
factor.  Second, after the end of contraction, and until the stabilization
point $b=b_0$ is obtained, both $a(t)$ and $b(t)$ expand.  Since
the normalcy bounds strictly only apply after stabilization is reached,
and the expansion of $a(t)$ red-shifts the temperature, the above
estimates of $T_{wall}$ are actually too high.

Having found reassurance in these numbers, we can ask if
this reheating could be not only indifferent, but beneficial
for late cosmology.  Namely, the bounds we have derived above suggest that
the reheating of the universe on the wall due to contraction may be
just sufficient to warm the world enough so that nucleosynthesis can occur
without hindrance. Indeed, the upper bound on the
reheat temperature is always
above the nucleosynthesis scale, suggesting a very interesting possibility
that in this picture reheating and particle production could be purely
gravitational. In conventional cosmological models based on quantum
gravity at the scale $10^{19}\gev$, gravitational particle production
is typically insufficient for reheating, largely due to the fact that
inflation occurs early, and that between it and the nucleosynthesis era
the universe has expanded by many orders of magnitude, diluting the
particles produced in the early de~Sitter phase.  In our model,
in contrast, the part of the world which contains our universe contracts,
increasing the energy density and number density on the wall rather
than diluting them.

\subsection{Bulk graviton production}

We close this section with the consideration of production rates for
bulk gravitons during the era of radion evolution.
There are actually two slightly different issues here that we have not
so far distinguished:  i) the energy density in the (would-be)
{\em zero mode} of the bulk graviton, namely the radion, and,
ii) the energy density in Kaluza-Klein (KK) excitations of the
graviton in the bulk.

The constraints on the energy density in KK excitations
is actually more severe than that on the radion energy
density, which just comes from overclosure.  The
reason for this is that even though the lifetime
of the bulk KK modes is very long, a small fraction of them will
decay back to photons on our brane, causing distortions in the
diffuse gamma-ray background ~\cite{ADDlong}.  The diffuse gamma-ray
background constraint only applies to the excited KK modes,
since in general it is only the excited states that can decay to
dangerous energetic SM states on our brane.  (Recall
that typically the mass of the radion, the would-be zero mode,
is $\gtap 10^{-3}\ev$.)  This constraint
is more severe than the overclosure constraint which applies to both
the radion zero mode and all the KK excitations together.

In any case in the next section we will consider the overclosure
constraint on the total bulk energy, and see that generically
there is a problem.  Here we want to check that the bound
on the KK excitations of the graviton is automatically satisfied.
(The analysis will show in passing that the dominant form
of bulk energy will be in the radion motion, and not the KK
excitations.)

First consider the evolution of the projected energy density
(of mass dimension $[\rho_{KK}] =4$ as usual)
of the bulk KK gravitons in the absence of particle production:
\be
\dot \rho_{KK} + 3H_a \rho_{KK} + H_b \rho_{KK} = 0 .
\ee
The second term on the LHS corresponds to the usual $1/a^3$
dilution of massive particles (which from this 4-dimensional
perspective the KK excitations of the graviton appear to be;
see \cite{ADDlong} for a full discussion).  The novel
feature of this equation is the third term which expresses
the fact that {\it as the internal radii increase, the KK masses
decrease as} $1/b$.  Another way of saying this is that the KK
mass is really the quantized momentum in the internal directions, and
this red-shifts as $b(t)$ expands.

As recognized long-ago by Parker and Zeldovich, and clarified later
by many authors~\cite{parker}, a time-dependent gravitational
field can produce particles from the vacuum, by essentially a
version of the pair-creation process that takes place in
strong electric fields for charged particles.  Up to order one
coefficients that will not be important, the magnitude of this
particle production per unit time per unit volume is given
by $H^5$, where $H$ is the typical Hubble constant.\footnote{More
precisely, as shown in \cite{parker}, in the anisotropic case considered
here, it is the Hubble constant of the contracting dimensions.}
Thus the equation for the projected KK energy density becomes
\be
\dot \rho_{KK} + 3H_a \rho_{KK} + H_b \rho_{KK} = H^5 .
\ee

It is easy to solve this equation. Using the substitution
$\rho_{KK} = f \Bigl(\frac{a_i}{a}\Bigr)^3 \Bigl(\frac{b_i}{b}\Bigr)$,
it can be reduced to
\be
\dot f = H^5 \left(\frac{a_i}{a}\right)^3
\left(\frac{b_i}{b}\right)
\ee
which, employing the Kasner era power laws, and noting that
the resulting power of $t$ is less than $-1$, gives upon integration
a result dominated by the {\it early} stages
\be
f \sim \frac{1}{t_i^4} - \frac{1}{t^4} \simeq H^4_i
\ee
for all power-law solutions.
Converted back into $\rho_{KK}$, this says
that the final energy density of the KK gravitons in the
bulk which came from particle production is bounded
from above by
\be
\rho_{KK,f} \ltap H^4_i \left(\frac{a_i}{a_f}\right)^3
\left(\frac{b_i}{b_f}\right).
\ee

This differs from our estimate of the final energy density
of the blue-shifted wall-localized radiation at the end of the
epoch of contraction, only in that it is further
suppressed by a factor of $(a_f/a_i)(b_i/b_f)$,
which comes from the fact that the KK gravitons are red-shifted
by the bulk expansion, but only diluted and not blue-shifted
by the wall contraction.  Thus we get
\bea
\frac{\rho_{KK,f}}{\rho_{wall,f}} &\le & \left(\frac{a_f}{a_i}\right)
\left(\frac{b_i}{b_f}\right)\\ \nonumber
& \ltap & \left(10^{1-30/n}\right)^{1 + \zeta},
\label{gravbound}
\eea
where in the second line we have used (\ref{contrrad}), and the
conservative estimates $b_i \sim 10 \mst^{-1}$ and $b_f=b_0$.
Evaluating this in, for example, the case of the simple Kasner
contraction with exponents given in (\ref{contrrad2}) leads
to $\rho_{KK}/\rho_{wall}$ varying between $3\times 10^{-17}$
for $n=2$ to $1\times10^{-8}$ for $n=6$.  This shows that
the effective temperature of the KK gravitons is well below
the diffuse gamma-ray bound, even before any dilution
necessary to solve the radion moduli problem.  It also
demonstrates that the vast majority of the energy in the
bulk is in the motion of the zero mode radion $\rho_b$, rather
than in the bulk KK modes.  This is simply because we have shown
above that $\rho_b\simeq \rho_{wall}$ is the natural circumstance
at the end of the epoch of contraction.

\subsection{The era of contraction and stabilization in the Einstein frame}

The epoch where the radius grows from its initial small size to
its final value can also be simply
and physically understood in the Einstein frame.
(Einstein frame quantities will be denoted by an overbar in the following.)
Recall that the metric in the Einstein frame is related to the
one in the string frame via
\be
\bar g_{\mu \nu} = e^{n \bar{\beta}} g_{\mu \nu}, ~~~~~~~~~~
\, b = e^{\bar{\beta}}
\ee
(throughout this subsection we work in units with $M_*=1$).
Using this map, it is straightforward to relate the quantities in
the string frame to their counterparts in the Einstein frame.
For instance, it is trivial to check that
\be
H_a = e^{\frac{n}{2} \bar{\beta}} \left(\bar H_a
- \frac{n}{2} \dot{\bar{\beta}} \right)
\ee
Therefore, we can see that even though in the Einstein frame
$\bar H_a >0$, this can look like ``contraction" in the
string frame, i.e. $H_a<0$, provided that $\dot{\bar{\beta}}$
is large enough. On the other hand, as long as the radius eventually
stabilizes at its final size so $\bar{\beta} = 0$, we
will have $H_a > 0$.  This proves that, in the string frame,
there is always a ``big bounce"
as long as the radius eventually stabilizes.

Moving on to the dynamics,
the action in the Einstein frame is
\be
S = \int d^4 y \sqrt{-\bar g} \left(- \bar {\cal R}  + \frac{n(n+2)}{2}
(\partial \bar{\beta})^2 - e^{-2 n \bar{\beta}} V(\bar{\beta}) \right)
\ee
We can parametrize the potential as
\be
V = b^{-p} W \equiv e^{-p \bar{\beta}} \bar{f}(\bar{\beta})
\ee
We wish to regard $\bar{f}$ as very slowly varying and will treat it as a
constant during the radion ``coasting". Of course, at
the final value of the radion
$\beta_*$ we must have
$\bar{f}(\beta) = \bar{f}'(\beta)= 0, \bar{f}''(\beta)>0$.

If we now work with the canonically normalized field
\be
\beta = \sqrt{n(n+2)} \bar{\beta}
\ee
the action becomes
\be
S = \int d^4 y \sqrt{-\bar g}  \left(- \bar {\cal R} + \frac12
(\partial \beta)^2 - e^{-x \beta} f(\beta) \right)
, ~~~~~~~~~~ \, x \equiv \frac{2n + p}{\sqrt{n(n+2)}}
\ee

The equations of motion are now just the familiar ones
for a FRW universe with a scalar field $\beta$ with an effective potential
$V_{eff}(\beta) = e^{-x \beta} f(\beta)$.  Writing the scale
factor as usual $a = e^{\alpha}$, we have
\bea
6 \dot{\alpha}^2 = \frac{1}{2} \dot{\beta}^2 + f e^{- x \beta} \\ \nonumber
\ddot{\beta} + 3 \dot{\alpha} \dot{\beta} = x f e^{-x \beta}
\eea
Suppose we ignore $f$; then $\beta$ only has kinetic energy and we
trivially find the solutions
\bea
\alpha = \frac{1}{3} \rm{log} (t) \nonumber \\
\beta = \frac{2}{\sqrt{3}} \rm{log} (t)
\eea
Obviously, for sufficiently large $x$, the effective potential
$V_{eff}$ is falling off so rapidly that ignoring $f$ should be a good
approximation. We can see what the lower bound on $x$ is by
substituting the above potential-free solutions into the exact equations
of motion; the kinetic energy terms scale like $t^{-2}$,
whereas the potential energy term scales like $t^{-2 x/\sqrt{3}}$.
Therefore, ignoring
the potential is a good approximation when
\be
x > \sqrt{3} \rightarrow x^2 - 3 = - \Delta >0
\ee
which agrees with the string frame result.  Notice
also that this case corresponds to
the string frame contraction, since
\be
{\rm sgn}(H_a) = {\rm sgn} \left(\dot{\alpha} -
\frac{n}{2 \sqrt{n(n+2)}} \dot{\beta}\right) =
{\rm sgn}\left(\frac{1}{3} - \frac{n}{\sqrt{3 n (n+2)}}\right) <0
\ee

What about $x < \sqrt{3}$ or $\Delta >0$? In this case, the
effective potential does not fall steeply enough for the energy density to
become kinetic energy dominated. What happens instead is that the
radion first gets accelerated by the potential till the kinetic
energy briefly dominates,
whereupon it gets diluted again and the cycle repeats. Therefore,
on average we expect the kinetic and potential energies to be
the same. It is easy to see that this assumption is self-consistently
justified by the equations. Therefore, we have
\be
\dot{\beta} = e^{-\frac{x}{2} \beta} \rightarrow
\beta = \frac{2}{x} \rm{log} (t)
\ee
Inserting this ansatz back into the equations, we find trivially
\be
\alpha = \frac{1}{x^2} {\rm log} (t)
\ee
These solutions correspond to the new, modified Kasner
solutions found in the string frame analysis.
Once again, it is easy to see that these solutions
correspond to contraction in the string frame:
\be
{\rm sgn}(H_a) = {\rm sgn}\left(\frac{1}{x^2} -
\frac{n}{\sqrt{n(n+2) x}}\right) = {\rm sgn}(2 - n - p)
\ee
while from our solutions in the string frame $a$ contracts when $n+p-2>0$.

We can now discuss what happens as the radion nears its final
minimum at $\beta=\beta_*$.
The effective potential $V_{eff}$
can be approximated as quadratic around this final point
for $(\beta - \beta_*)/\beta_* \lsim 1$. If the radion approaches
this region with kinetic energy sufficiently
smaller than, or comparable to, the potential energy,
then by equipartition it will be trapped in the well and
will oscillate about the minimum.
On the other hand, if the kinetic energy is much larger than
the potential at the top of the well,
$\beta$ will escape from the region close to the minimum;
its subsequent fate depends on the form of $V_{eff}$ at
larger values of $\beta$.
It can either escape to infinity,  or it may turn back around
at some large distance away and ``slosh" back and forth with
very large amplitude about
$\beta_*$.  It is therefore clear that for $x <\sqrt{3}$, where
the kinetic and potential energy stay comparable throughout the
coasting period,
the radion will not overshoot. On the other hand, it appears that
for $x > \sqrt{3}$, where the kinetic energy dominates, the
radion could significantly overshoot the minimum.
This statement has to be qualified, however, since all of the
above analysis neglects the effect of the radiation energy density
left over after inflation.
Of course, in the Einstein frame,this energy does not appear to be
blue-shifted, because the Einstein frame scale factor never
decreases.
However, during the kinetic energy dominated era,
the kinetic energy redshifts as $\bar{a}^{-6}$ whereas the radiation energy density only
redshifts like $\bar{a}^{-4}$,  so the radiation will eventually dominate.
Let us consider what happens after radiation domination, ignoring the radion potential.
Since $\bar{a} \sim \sqrt{t}$ during radiation domination, the radion kinetic energy $\dot{\beta}^2 \sim
\bar{a}^{-6} \sim t^{-3}$, so $\dot{\beta} \sim t^{-3/2}$ giving $\beta \sim t^{-1/2}$. In other words, ignoring the radion
potential, the radiation provides enough friction to stop $\beta$. Therefore, if radiation domination happens {\it before}
$\beta$ approaches the minimum $\beta_*$, radiation domination prevents overshoot.
In all cases of interest to us this indeed happens; as demonstrated numerically in
subsection 3.2, radiation domination takes place for values of the radius $b$ less than the minimum $b_0$.

\subsection{Stabilization and the moduli problem}

In the previous subsections we discussed a rather interesting
mechanism by which the remnant Hawking radiation left over
from the de~Sitter era could be the source of reheating on our
brane when combined with the fact that $a(t)$ goes through a
period of contraction.

However as we noted above, before this mechanism can be viewed
as a realistic way to reheat our brane, we must consider the
radion field.  At the stage where the
contraction stops and the brane undergoes a Big Bounce, the contribution
of the energy density of the radion field is still a significant
contribution to the total (indeed $O(1)$).
This fact leads to a radion moduli problem. The difficulty is
that the radion is so light and weakly coupled that it
lives typically much longer than the age of the universe. Since
its coherent oscillations about the minimum redshift away only as $1/a^3$,
it can eventually overclose universe.  In order to avoid this, the
energy stored in the radion must be small, relative to the
radiation energy $T_*^4$, when the universe is reheated to
$T_*$:
\be
\frac{\rho_{rad *}}{T_*^4} = \frac{\rho_{rad *}}{T_*^3}
\times \frac{1}{T_*} = \frac{\rho_{rad 0}}{T_0^3}
\times \frac{1}{T_*} < \frac{3 \times 10^{-9}\gev}{T_*}
\ee
where we have used $\rho/T^3 = {\rm const.}$ and
$\rho_{crit 0}/T_0^3 \sim 3 \times 10^{-9}\gev$.
Given that $T_*$ is bounded between $\sim 1$ MeV - 1 GeV by normalcy constraints,
the energy density in the radion must somehow be diluted
by $\sim 10^{-7} - 10^{-9}$ in order to avoid overclosure.
Fortunately this implies that, for example, only about 5 or 6
efolds of late inflation are needed so solve this problem.
In any case this moduli problem is one of late (post-stabilization)
cosmology, rather than early pre-stabilization cosmology;
we will return to address it in a future publication.

\section{Conclusions}

We have argued that early inflation when the internal
dimensions are still small can successfully accomplish
all that is required of inflation, including generation
of suitable $\de\rho/\rho$ without the unpleasant introduction
of very light or fine-tuned wall fields.
Indeed, the very fact that the internal dimensions must expand from
their initial size close to the fundamental Planck length $\mst^{-1}$
to the stabilization value $b_0 \simeq 10^{-17 + 30/n} {\rm cm}$
leads to natural inflationary scenarios
involving the dynamics of the internal space.  The resolution
of cosmological conundrums such as the horizon, flatness and age problems,
and the production of the spectrum of nearly-scale invariant
Harrison-Zeldovich density perturbations with the avoidance of drastic
fine tuning of the inflaton mass come as a consequence of the evolution
of the internal space.  Moreover it is remarkable that the era of
post-inflation brane-contraction that follows this period of
inflation is harmless, and automatically ends via a ``Big Bounce''.
During the phase of $b(t)$ evolution to the
stabilization point, the production of bulk
gravitons by the time-varying metric remains completely suppressed,
ensuring that the bulk is very cold at, and after, the stabilization
of the internal dimensions.  The primary remaining issue is the radion
moduli problem, which is no more severe than in gauge-mediated
supersymmetry breaking models.  Overall, then, early universe cosmology
in these models is quite interesting!

\section*{Acknowledgments}
It is a pleasure to thank Gia Dvali for valuable discussions of related
ideas \cite{GIA} and to Andrei Linde for useful conversations.
SD thanks the CERN theory group, and JMR thanks the
Stanford University theory group, for their respective hospitality
during portions of this work.  The work of SD and NK is supported in
part by NSF grant PHY-9870115. The work of JMR is supported
in part by an A.P. Sloan Foundation Fellowship.

\section*{Appendix A: Kinematics of the radion field}

With the metric of the form (\ref{metricform}) the Ricci scalar is
\be
-{\cal R} = 6{ \ddot a\over a } + 6\left({ \dot a\over a}\right)^2
+ 2n{ \ddot b\over b } + n(n-1)\left({ \dot b\over b }\right)^2 +
6n\left({ \dot a\dot b\over ab }\right) + {\kappa n(n-1) \over
b^2},
\label{ricci}
\ee
where the internal curvature term is present for $n$-spheres
($\kappa=1$), but vanishes for tori ($\kappa=0$), and we have ignored
a similar curvature term for the large dimensions.  After integrating
over all spatial coordinates we obtain,
\be
S = \int dt \bigl( {\cal L}_{\rm KE}(\dot a, \dot b) - a^3 V(b) \bigr),
\label{intaction}
\ee
and further integrating the $\ddot a$ and $\ddot b$ terms by parts, the
kinetic part of the action for the radii, $a$ and $b$, becomes
\be
S = -\mst^{2+n}\int dt\, a^3 b^n \left( 6\left({ \dot a\over a
}\right)^2 +  n(n-1)\left({ \dot b\over b }\right)^2 + 6n\left({
\dot a\dot b\over ab }\right) \right) .
\label{radiiKE}
\ee
Note the overall negative sign of these kinetic terms, but also
the mixing between $\dot a$ and $\dot b$.

There is clearly an extremum of the action with $\dot
a =\dot b=0$, when the condition $\partial_a ( a^3 V_{\rm
tot}(b))|_{a=a_0, b=b_0}=0$, and similar with
$\partial_a\to\partial_b$ are met.  These imply (for $a_0\neq0$)
\bea
V_{\rm tot}(b_0) & = & 0, \qquad {\rm and}\cr V'_{\rm tot}(b_0) &=
& 0 .
\label{staticcond}
\eea
This is as one would have naively expected. However, because of
the negative sign for the kinetic term for the radial degrees of
freedom, the stability analysis for such static solutions has to
be treated with care.  The analysis starts by expanding the
action, Eq.~(\ref{radiiKE}), in small fluctuations around the
extremum: $a(t) = a_0 +\delta a(t)$, and $b(t) = b_0 +\delta
b(t)$. Then to quadratic order, and defining $\Delta \equiv \delta
a/a_0$ and $\delta \equiv \delta b/b_0$, the expansion gives the
coupled equations of motion
\be
\left(
\begin{array}{cc}
6 & 3n \\ 3n & n(n-1)
\end{array}\right)
\left(
\begin{array}{c}
\ddot \Delta \\
\ddot \delta
\end{array}\right)
=
\left(
\begin{array}{cc}
0 & 0 \\
0 & \omega^2
\end{array}\right)
\left(
\begin{array}{c}
\Delta \\
\delta
\end{array}\right),
\label{stabM}
\ee
where
\be
\omega^2 = \frac{1}{2} { (b_0)^2 V''_{\rm tot}(b_0)\over
\mst^{2+n}(b_0)^n} = \frac{1}{2} { (b_0)^2 V''_{\rm tot}(b_0)\over
\mpl^{2}}.
\label{omegadef}
\ee
is the radion mass around the stabilization point.
Searching for oscillating solutions, $(\Delta,\delta) = \exp(i\Omega
t)(\Delta_0,\delta_0)$ of the stability equations leads
to an eigenvalue problem for the frequency $\Omega$.
Specifically, $\Omega^2$ has the eigenvalues
$\Omega^2 =0$, and
\be
\Omega^2 = {2\over n(2+n)}\omega^2.
\label{Omegaeqn}
\ee

The zero eigenvalue just corresponds to the fact that $a_0$ is a
flat direction since, by assumption, there is no potential for
$a$. The crucial expression is Eq.~(\ref{Omegaeqn}), which gives
the condition for stability of the static solution. In the end,
stability just requires that the radion (mass)$^2$ be positive
as one would expect, and that we can think in
terms of a total potential $V(b)$ that one can minimize to find
the stable static solutions for the size of the internal
dimensions.

The equations of motion for $a(t)$ and $b(t)$ derived from the
action Eqs.~(\ref{intaction}) and (\ref{radiiKE}) are, after some
algebra, those given in the text, namely (\ref{aeom}), (\ref{beom})
and the constraint (\ref{constraint}) which comes about from the
well-known property that the total energy in GR is zero. (This
constraint can be derived by carefully working in terms of the
lapse and shift functions of the canonical formalism.) If matter
on our wall is also included then they become those given
in (\ref{eoms}):
\bea
&&6 H_a^2 + n(n-1) H_b^2 + 6n H_a H_b = \frac{V + \rho}{\mst^{n+2} b^n}
\nonumber \\
&&\frac{\ddot b}{b} + (n-1) H_b^2 + 3 H_b H_a
= \frac{1}{\mst^{n+2} b^n} \left(\frac{2 V}{n+2}
- \frac{b}{n(n+2)} \frac{\partial V}{\partial b} +
\frac{\rho-3p}{2(n+2)} \right) \nonumber \\
&& \frac{\ddot a}{a} + 2H_a^2 + n H_b H_a = \frac{1}{\mst^{n+2} b^n} \left(
\frac{b}{2(n+2)} \frac{\partial V}{\partial b}
- \frac{n-2}{2(n+2)} V +
\frac{\rho + (n-1) p}{2(n+2)} \right) \nonumber  \\
&&\dot \rho + 3H_a(p + \rho) = 0.
\eea
Note that in these equations the effect of wall-localized matter
is just some extra contribution to the $a(t)$ and $b(t)$ scale-factor
evolution.  Of course the energy density on the brane in general
distorts the {\it geometry} of the internal space (as does that
present on other branes that may exist in the bulk), but as far
as the overall properties and evolution of the {\it zero mode
size modulus} $b(t)$ of the internal space is concerned, it is correct
to treat the energy density on the wall as just {\it averaged}
over the whole space, as done on the RHS of these equations.

It is useful to summarize some basic properties of the evolution
equations.  Since the total potential energy $U(a,b,\psi)$ is
given in terms of the effective 4-dimensional potential energy
density $V(b,\psi)$ by $U = a^3 V(b,\psi)$, a uniform bulk
cosmological constant is represented by $V= b^n \Lambda$.
Substituting this form into the
RHS of these equations we see that a positive bulk
cosmological constant term has the effect of wanting to increase
{\it both} $\ddot a$ {\it and} $\ddot b$ if the evolution of $a,b$
are studied close to zero.  This is not inconsistent with the
stability criteria derived above since
this was explicitly the stability analysis around a stationary
point of the equations with non-zero values of both $a$ and $b$.
Indeed, the stability analysis can be derived directly from the
equations of motion, as it
must.  Concretely, if we expand around a point $(a_*,b_*)$ with
$V(b_*)=V'(b_*)=0$, then linearizing the equations of motion gives
\be
\left(
\begin{array}{c}
\delta \ddot a \\ \delta \ddot b
\end{array}\right)
=
\left(
\begin{array}{cc}
0 & \frac{a_* b_*^{1-n}}{2(n+2)\mst^{n+2}} V'' \\ 0 &
-\frac{b_*^{2-n}}{n(n+2)\mst^{n+2}} V''
\end{array}\right)
\left(
\begin{array}{c}
\delta a \\ \delta b
\end{array}\right),
\label{stabeom}
\ee
which exactly reproduces the previous stability analysis, in
particular the requirement $V''(b_*)>0$ for $a_*,b_* > 0$.

Another basic property that is useful to keep in mind is the effect
of some small amount of wall-localized matter on the position and
mass of the radion.  Linearizing the equations (\ref{eoms}) with
such matter in the shift $\de_b =\de b/b_0$ around the stabilization point
we find
\be
\ddot\de_b +3\dot\de_b H_a = -\frac{1}{(n+2) b_0^n \mst^{n+2}}
\left(\frac{b_0^2 V''}{n} + \frac{n(\rho-3p)}{2}\right) \de_b +
\frac{(\rho-3p)}{2(n+2) b_0^n \mst^{n+2}}.
\ee
This shows that wall-localized matter has two effects o the
radion.  First, the $\de_b$-independent term on the RHS shifts
the radion from the stabilization point by a small amount, and second
the $(\rho,p)$-dependent in the parentheses on the RHS shifts
the eigenfrequency of oscillations of the radion (or equivalently
the radion mass).  Both these effects are to be expected, and are
harmless for wall-localized matter densities and pressures
$\rho,p\ll \mst^4$.

\section*{Appendix B: Kasner-like solutions with potential}

The solutions of (\ref{lasteq}) take different
form, controlled by whether $C_2$ vanishes or not
and the sign of $\Delta$. Let us first consider
the case $\Delta > 0$. Then, when $C_2 = 0$, the equation
(\ref{lasteq}) simplifies to
$X' = \pm \sqrt{2 \frac{\Delta}{n(n+2)} \omega} \exp(X/2)$,
which can be integrated to give $\exp(X)
= \frac{2n(n+2)}{\Delta \omega} \frac{1}{\tau^2}$.
Here we have removed an additional integration constant
by a time translation. It is straightforward now to use
this and (\ref{oneintegr}) to determine the solutions for $\alpha$ and
$\beta$. Since
$\alpha = \frac{(2n+p)(n-p) C_1}{4\Delta} - \frac{n(n+p-2)}{2\Delta}X$ and
$\beta = \frac{2n+p}{\Delta} X - \frac{3(2n+p) C_1}{2\Delta}$,
we find
\bea
e^{\alpha} &=& \exp\left(\frac{(2n+p)(n-p) C_1}{4\Delta}\right)
\left(\frac{\Delta \omega}{2n(n+2)}\right)^{\frac{n(n+p-2)}{2\Delta}}
\tau^{\frac{n(n+p-2)}{\Delta}}
\nonumber \\
e^{\beta} &=& \exp\left(-\frac{3(2n+p)C_1}{2\Delta}\right)
\left(\frac{2n(n+2)}{\Delta \omega}\right)^{\frac{2n+p}{\Delta}}
\frac1{\tau^{\frac{4n+2p}{\Delta}}} .
\label{solsone}
\eea
We can now use (\ref{newtime}) to transform (\ref{solsone})
back to using the comoving time on the brane. Integrating
(\ref{newtime}), we find
\be
\tau \sim t^{-\frac{\Delta}{(n+p)(2n+p)}}.
\label{onetime}
\ee
Note the important feature of this map, that it maps the
comoving future $t \rightarrow \infty$ to the origin of
``time" $\tau$ and vice-versa.  Thus to discover the correct long-time
asymptotics of the solutions we must extract the small-$\tau$
behavior.  This is true for the general case $C_2 \ne 0$ as well.

In any case, for $C_2=0$, and after appropriate rescalings,
we find the behavior of the scale factors $a(t)$ and $b(t)$
to be
\bea
a &=& a_i \left(\frac{t}{t_i}\right)^{-\frac{n(n+p-2)}{(n+p)(2n+p)}}
\nonumber \\
b &=& b_i \left(\frac{t}{t_i} \right)^{\frac{2}{n+p}} .
\label{solsonefin2}
\eea
These solutions describe the cases where the brane length scales
shrink while the radion continues to grow.  This means that
even if the brane scale factor $a(t)$ continues to
expand by inertia after the end of inflation, soon after
$a(t)$ starts to contract.  Moreover, the behavior (\ref{solsonefin2})
is an asymptotic attractor for the generic cases with $C_2<0$ and
$\Delta > 0$, as we will now discuss.

Let us now consider the case when $C_2 < 0$.
Again, note from (\ref{oneintegr}) that this corresponds
to the situation where initially both $H_a$ and $H_b$ are
positive, when inflation has just ended. The equation (\ref{lasteq})
can be solved in this case by the substitution
\be
e^{-X} = \frac{8\Delta \omega}{3(2n+p)^2 C^2_2} \sinh^2(\vartheta)
\label{subs}
\ee
which reduces to the differential equation $
\vartheta' = \Bigl(\frac{3(2n+p)^2}{16n(n+2)}\Bigr)^{1/2} C_2$, whose
solution is $\vartheta =
\Bigl(\frac{3(2n+p)^2}{16n(n+2)}\Bigr)^{1/2} C_2 \tau$,
after the appropriate choice of the coordinate origin of $\tau$.
Therefore, the solution of (\ref{lasteq}) is
\be
e^{-X} = \frac{8\Delta \omega}{3(2n+p)^2 C^2_2}
\sinh^2(\sqrt{\frac{3(2n+p)^2}{16n(n+2)}} C_2 \tau)
\label{solc}
\ee
Note that the RHS is a positive semidefinite function,
guaranteeing the reality of the metric, as required.
Also note that the solution (\ref{solc}) is defined
in open interval $\tau \in (0^+,\infty)$
and $(-\infty, 0^-)$, by time-reversal.
Using this equation and (\ref{oneintegr}),  we
can extract the solutions for $\alpha$ and $\beta$:
\bea
e^{\alpha} &=& \Bigl(\frac{8\Delta \omega}{3(2n+p)^2 C^2_2}
\Bigr)^{\frac{n(n+p-2)}{\Delta}}
e^{\frac{(2n+p)(n-p)}{4\Delta}(C_1 + C_2 \tau)}
\sinh^{\frac{n(n+p-2)}{\Delta}}(\sqrt{\frac{3(2n+p)^2}{16n(n+2)}}
|C_2 \tau|)
\nonumber \\
e^{\beta} &=& \Bigl(\frac{3(2n+p)^2 C^2_2}{8\Delta \omega}
\Bigr)^{\frac{2n+p}{\Delta}} \frac{\exp\Bigl(-\frac{3(2n+p)}{2\Delta}
(C_1 + C_2 \tau)\Bigr)}{\sinh^{2\frac{2n+p}{\Delta}}
(\sqrt{\frac{3(2n+p)^2}{16n(n+2)}} |C_2 \tau|)}
\label{solsc}
\eea
where the argument of the hyperbolic sine is taken
to be positive, to make sure that the solutions remain
real.  In general, it is not possible to explicitly determine
the integrated form of the gauge transformation (\ref{newtime}).
However, the asymptotic limits $\tau \rightarrow 0, \infty$
are easy to deduce, and are sufficient for our purpose here.

Taking the $\tau\to 0$ limits of (\ref{solsc}) we find that
(\ref{solsc}) reduce precisely to (\ref{solsone}), implying that
the integrated form of (\ref{newtime}) approaches (\ref{onetime}).
Hence in the case $C_2<0$ the scale factors again approach
(\ref{solsonefin}) as $t \rightarrow \infty$, which are therefore
the appropriate future attractors in all cases
$\Delta>0$, $C_2\le 0$.

It is also amusing to consider the short $t$-time
behavior of the exact solutions in this case.  We
find that as $\tau \rightarrow \infty$,
\bea
e^\alpha &\sim& \exp\Bigl(\frac{(2n+p)C_2 \tau}{4\Delta}
(n-p - n(n+p-2)\sqrt{\frac3{n(n+2)}}) \Bigr) \nonumber \\
e^\beta &\sim& \exp\Bigl(-\frac{(2n+p)C_2 \tau}{2\Delta}
(3 - (2n+p)\sqrt{\frac3{n(n+2)}}) \Bigr)
\label{asympsol}
\eea
Using this, we can integrate (\ref{newtime}) and find
\be
t \sim \exp\Bigl(\frac{(2n+p)C_2 \tau}{4\Delta}
(n(n-p+6)\sqrt{\frac3{n(n+2)}} - 3(n+p)) \Bigr)
\label{transf}
\ee
and upon substituting this back into (\ref{asympsol}),
we can show that the solutions approach from below
the asymptotic expressions
\bea
e^\alpha &\rightarrow&
\Bigl(\frac{t}{t_i}\Bigr)^{\frac{3+\sqrt{3n(n+2}}{3(n+3)}}
\nonumber \\
e^\beta &\rightarrow&
\Bigl(\frac{t}{t_i}\Bigr)^{\frac{n-\sqrt{3n(n+2)}}{n(n+3)}}
\label{asympsollast}
\eea
The dependence on the parameter $p$ apparent in
(\ref{asympsol}) has completely disappeared. This can be most easily
seen by taking the derivatives with respect to $p$ of the powers
which are obtained by substituting (\ref{transf}) into (\ref{asympsol}),
noting that they are identically zero,
and then setting $p=0$ to obtain (\ref{asympsollast}).
Also note that the resulting powers are identical to those found
in the simple Kasner case (\ref{kasnercoeff}), with lower sign taken.
Thus in the very early time limit, this suggests that the universe
was expanding while the radion was decreasing. However, this phase
is cut out by a stage of inflation, and in fact only a very short
portion is retained where both $a$ and $b$ are growing for short time,
to match onto the post-inflationary era.

Now we consider the last case, $\Delta < 0$, which
we argued in the text should be described by the traditional
potential-free Kasner solutions in the long-time limit.
We can see this explicitly by examining the exact solutions
once again.  First note that, in this case, $C_2$ must
be nonzero, as can be immediately seen from (\ref{lasteq}).
The exact solutions can be found by the substitution
\be
e^{-X} = \frac{8|\Delta| \omega}{3(2n+p)^2 C^2_2} \cosh^2(\vartheta)
\label{subsc}
\ee
which is analogous to (\ref{subs}). The equation for $\vartheta$
is the same as before, and hence the solution is
\be
e^{-X} = \frac{8|\Delta| \omega}{3(2n+p)^2 C^2_2}
\cosh^2\left(\sqrt{\frac{3(2n+p)^2}{16n(n+2)}} C_2 \tau\right)
\label{solcc}
\ee
The similarity of this solution to (\ref{solc}) allows us to
extract the expressions for $\alpha$ and $\beta$ quite easily. We
find
\bea
e^{\alpha} &=& \Bigl(\frac{8|\Delta| \omega}{3(2n+p)^2 C^2_2}
\Bigr)^{\frac{n(n+p-2)}{|\Delta|}}
e^{\frac{(2n+p)(n-p)}{4|\Delta|}(C_1 + C_2 \tau)}
\cosh^{\frac{n(n+p-2)}{|\Delta|}}\left(\sqrt{\frac{3(2n+p)^2}{16n(n+2)}}
|C_2 \tau|\right)
\nonumber \\
e^{\beta} &=& \Bigl(\frac{3(2n+p)^2 C^2_2}{8|\Delta| \omega}
\Bigr)^{\frac{2n+p}{|\Delta|}} \frac{\exp\Bigl(-\frac{3(2n+p)}{2|\Delta|}
(C_1 + C_2 \tau)\Bigr)}{\cosh^{2\frac{2n+p}{|\Delta|}}
\left(\sqrt{\frac{3(2n+p)^2}{16n(n+2)}} |C_2 \tau|\right)}
\label{solscc}
\eea
Note however the important difference between these
solutions and (\ref{solsc}).
In this case, the solutions are defined on the
whole interval $(-\infty, \infty)$.
By considering the limits $\tau \rightarrow \pm \infty$, we can see
that the solutions (\ref{solscc}) in fact interpolate between the
simple Kasner solutions (\ref{kasner}), where for $C_2 < 0$ it
starts out with powers given by (\ref{kasnercoeff}) with lower
sign and transmutes due to the intermediate potential-dominated region
into (\ref{kasner}) with powers (\ref{kasnercoeff}) with upper sign
taken. In reality, the far past of these solutions is cut out by the
inflationary era, and again only a very short portion where both $a$
and $b$ are growing is retained, which is matched
onto the post-inflationary era.
All of these solutions therefore flow towards simple Kasner solutions
with the powers with upper sign in (\ref{kasnercoeff}), which are the
appropriate future attractors.

\section*{Appendix C: Exact Solutions for the Big Bounce}

To see that bounce behavior does indeed occur in the presence
of radiation on the wall we go back to our equations of motion
but now set the potential to zero and just keep the wall-localized
radiation terms on the RHS:
\bea
&&6 H_a^2 + n(n-1) H_b^2 + 6n H_a H_b = \frac{ \rho}{\mst^{n+2} b^n}
\nonumber \\
&&\frac{\ddot b}{b} + (n-1) H_b^2 + 3 H_b H_a
= \frac{\rho-3p}{2(n+2)\mst^{n+2} b^n} \nonumber \\
&& \frac{\ddot a}{a} + 2H_a^2 + n H_b H_a = \frac{1}{\mst^{n+2} b^n}
\frac{\rho + (n-1) p}{2(n+2)} \nonumber  \\
&&\dot \rho + 3H_a(p + \rho) = 0.
\eea
These can again be solved exactly, and provide a good approximation
for the exit from the Kasner-like phase of $a(t)$-contraction.

Set $p=\rho/3$ for radiation.  The solution for the
radiation energy density is as usual
\be
\rho = \frac{B \mst^{n+2}}{a^4}
\ee
(with some normalization $B$) and
define variables $\alpha$ and $\beta$ by
\bea
a &=& a_0 e^{\alpha} \nonumber \\
b &=& b_0 e^{\beta}
\eea
where $a_0$ and $b_0$ are constants, determined by the values
of $a$ and $b$ at the end epoch when the radion potential
becomes small compared to the wall-localized radiation.  Using
this, the equations of motion can be rewritten as
\bea
&&6 \dot \alpha^2 + n(n-1) \dot \beta^2 + 6n \dot \alpha \dot \beta
= Ae^{-n \beta-4\alpha} \nonumber \\
&&\ddot \beta + \dot \beta(3\dot \alpha + n \dot \beta)
= 0
\nonumber \\
&& \ddot \alpha + \dot \alpha( 3\dot \alpha + n \dot \beta)
= \frac{A}{6} e^{-4\alpha-n \beta}
\eea
where $A = B/(a^4_0 b^n_0)$. If we
change the time variable, defining the new time $\tau$ as
in eq. (\ref{newtime}):
\be
d\tau = -e^{-3\alpha - n \beta} dt
\ee
where again the
specific form of $\tau = \tau (t)$ can be found later
when solutions are determined,
we can rewrite the equations of motion as
\bea
&&6 \alpha'^2 + n(n-1) \beta'^2 + 6n \alpha' \beta'
= A e^{2\alpha + n \beta} \nonumber \\
&&\beta'' = 0
\nonumber \\
&& \alpha'' = \frac{A}{6} e^{2\alpha +n \beta}
\eea
where the primes denote derivatives with respect
to $\tau$. The $\beta$ equation immediately gives
\be
\beta = C_1 + C_2 \tau
\ee
Hence note that if $H_b>0$ initially then the constant $C_2<0$.
Also note that given our conventions we again
have future $t$ infinity map to $\tau = 0$.

For the purpose of examining the solutions to these
equations we distinguish two cases:

\begin{itemize}

\item[(i)] $C_2 = 0$; like before, this solution turns out
{\it to be a future attractor of all solutions}!

\end{itemize}

This case has a very simple analysis: $C_2=0$ implies
$\beta'=0$ but also $\dot \beta = 0$. Hence we can forget about
the $\tau$ coordinates and immediately work in our original
brane-time $t$. So $\dot b =0$ means $b=b_0 = const$ and so
the constraint equation gives
\be
6 H_a^2 = \frac{A}{b_0^n a^4}
\ee
which is immediately solved to give us the radiation
dominated universe, $a \sim \sqrt{t}$.  This is precisely
as we would expect from an analysis in the Einstein frame.

\begin{itemize}

\item[(ii)] $C_2 \ne 0$; this case corresponds to more generic solutions.
In particular we will see that we can have initial conditions
$H_b>0$, $H_a<0$, which evolve to $H_b=0$, $H_a>0$, in other words
a ``bounce'' solution.

\end{itemize}

To solve the equations in this case, first define
\be
X = 2 \alpha + n \beta,
\ee
in terms of which, the second order differential equations produce
\be
X'' = \frac{A}{3} e^X
\ee
This is again the Liouville equation, with the first
integral (conservation of energy)
\be
X'^2 = \frac{2A}{3} e^X + Z_0
\ee
where $Z_0$ is another integration constant.
If we
eliminate $e^X$ from the constraint equation by using the last equation,
and then simplify the result by using the second equation,
we find that the constraint reduces to
\be
Z_0 = \frac{n(n+2)}{3} C^2_2 \ge 0
\ee
Thus the constant $C_2$ controls the dynamics, and the whole
system has collapsed down to two simple equations,
one of which is already solved:
\bea
&&\beta = C_1 + C_2 \tau
\nonumber \\
&&X'^2 = \frac{2A}{3} e^X + \frac{n(n+2)}{3} C^2_2 .
\eea

Note that the solution to the $X$ equation is,
\be
e^X = \frac{n(n+2)C^2_2}{2A}
\frac{1}{\sinh^2(\sqrt{\frac{n(n+2)}{12}} C_2 \tau)}
\ee
as may be checked by substitution.
Using the formula for $X=2\alpha + n\beta$ and
$\beta = C_1 + C_2 \tau$, we find that
\be
e^{\alpha} = +\left(\frac{n(n+2) C^2_2}{2A}\right)^{1/2}
\frac{e^{-nC_1/2 + n|C_2|\tau/2}}{\sinh(\sqrt{\frac{n(n+2)}{12}} |C_2| \tau)},
\label{bouncesol}
\ee
where since $C_2<0$ in the cases of interest we have taken the
appropriate branch of the square-root such that $a=a_0 e^\alpha$
is positive as it must be.

This expression is already sufficient to show that we get a
bounce behavior for $a(t)$.  Recognizing that $t=0$ corresponds
to $\tau\to\infty$ while $t\to\infty$ corresponds to $\tau\to 0$,
simply plotting $e^\alpha$ as given by (\ref{bouncesol})
shows that $a(t)$ goes through a bounce.

It is instructive to see this in detail.  First,
consider the limit $\tau \rightarrow 0$. Clearly,
$\beta =C_1 + C_2 \tau \rightarrow C_1 = const$. Also,
\be
e^{\alpha} \rightarrow + \left(\frac{6e^{-nC_1}}{A}\right)^{1/2}\frac{1}{\tau}
\ee
Thus $\exp(3\alpha + n\beta) \rightarrow \tau^{-3}$, and so we have
\be
dt \sim -\frac{d\tau}{\tau^3}
\ee
or, $t \sim \tau^{-2}$ which gives us $\tau \sim 1/\sqrt{t}$, and
$t\to \infty$ maps to $\tau \to 0^+$ as claimed.
Back in the formula for $a$, this gives
\be
a \sim \exp(\alpha) \sim \frac{1}{\tau} \sim \sqrt{t}
\ee
and this is precisely a radiation-dominated universe
at late times!  So indeed, the solutions with
$C_2 = 0$ are late time attractors.
Some further analysis shows that in this limit, (and
in the approximation of ignoring the stabilizing potential)
$b$ will tend to a constant logarithmically.
The end result of the analysis is that for large $t$
these solutions show that $a$ is expanding asymptotically as
$\sqrt{t}$.

Let's now look at the other limit, $\tau \rightarrow \infty$.
Also, recall that $C_2$ must be less than zero: $C_2 < 0$.
We have
\be
b \sim \exp(\beta) \sim \exp(-|C_2| \tau)
\ee
and
\be
a \sim \exp(\alpha) \sim
\exp\left(\frac{|C_2|\tau}{2}\Bigl(n-\sqrt{n(n+2)/3}\Bigr)\right).
\label{atautoinfty}
\ee
These give the relationship between $\tau$ and $t$
\be
dt \sim - \exp\left(\frac{|C_2|\tau}{2}\Bigl(n-\sqrt{3n(n+2)}\Bigr)\right)
d\tau
\label{tautoinfty}
\ee
Now note that $(n-\sqrt{3n(n+2)}) < 0$ for all $2\le n\le 6$, and
therefore as $\tau\to +\infty$ we have $t\to 0^+$ as
\be
t \sim \exp\left(\frac{-|C_2|\tau}{2}(\sqrt{3n(n+2)}-n)\right)
\ee
as claimed.  Finally, in the expression (\ref{atautoinfty}) for
$a(t)$, note that $(n-\sqrt{n(n+2)/3}) > 0$, so as
$\tau\to \infty$, the scale factor $a(t)$ is again going to
infinity.  Re-expressed in terms of $t$ we see that as $t$ initially
increases $a(t)$ is initially {\it decreasing}.

If one carefully considers this limit the
power law behavior of $a$ and $b$ in terms of the time
$t$ precisely corresponds to the late-time Kasner solutions
we found in Appendix~B.  Most importantly,
we see that as $\tau$ decreases from $\infty$
towards $0^+$, the radion $b$ is initially increasing and
$a$ is initially decreasing, the initial conditions
that we require.
But as we have seen from the analysis in the $\tau\to 0$
limit, this goes over to
$b$ growing logarithmically and $a$ increasing as $1/\tau$.
This implies there must indeed have been a Big Bounce in between!

Finally, the bounce occurs when $\dot \alpha = 0$ or equivalently
$H_a=0$.  This happens when
\be
n(n-1) H^2_b = \frac{\rho_{wall}}{\mst^{n+2} b^n}.
\ee
Hence, when the bounce occurs the radion kinetic energy
is comparable to the wall energy density. By continuity,
it should be clear that the generic qualitative features of these
properties would remain true even in the presence of stabilizing
potentials. The main conclusion of this analysis is that
the Big Bounce is the future asymptotic attractor of all
postinflationary solutions with wall radiation, and hence
the exit from the contraction on the wall will occur
naturally.

\def\pl#1#2#3{{\it Phys. Lett. }{\bf B#1~}(19#2)~#3}
\def\zp#1#2#3{{\it Z. Phys. }{\bf C#1~}(19#2)~#3}
\def\prl#1#2#3{{\it Phys. Rev. Lett. }{\bf #1~}(19#2)~#3}
\def\rmp#1#2#3{{\it Rev. Mod. Phys. }{\bf #1~}(19#2)~#3}
\def\prep#1#2#3{{\it Phys. Rep. }{\bf #1~}(19#2)~#3}
\def\pr#1#2#3{{\it Phys. Rev. }{\bf D#1~}(19#2)~#3}
\def\np#1#2#3{{\it Nucl. Phys. }{\bf B#1~}(19#2)~#3}
\def\mpl#1#2#3{{\it Mod. Phys. Lett. }{\bf #1~}(19#2)~#3}
\def\arnps#1#2#3{{\it Annu. Rev. Nucl. Part. Sci. }{\bf #1~}(19#2)~#3}
\def\sjnp#1#2#3{{\it Sov. J. Nucl. Phys. }{\bf #1~}(19#2)~#3}
\def\jetp#1#2#3{{\it JETP Lett. }{\bf #1~}(19#2)~#3}
\def\app#1#2#3{{\it Acta Phys. Polon. }{\bf #1~}(19#2)~#3}
\def\rnc#1#2#3{{\it Riv. Nuovo Cim. }{\bf #1~}(19#2)~#3}
\def\ap#1#2#3{{\it Ann. Phys. }{\bf #1~}(19#2)~#3}
\def\ptp#1#2#3{{\it Prog. Theor. Phys. }{\bf #1~}(19#2)~#3}

\end{document}